\documentclass[longauth]{aa}
\usepackage[caption=false]{subfig} 
\usepackage{graphicx}
\usepackage{xcolor}
\usepackage{enumitem}
\usepackage{multirow}
\usepackage{float} 
\usepackage{txfonts}
\usepackage{natbib,twoopt}
\usepackage[breaklinks=true]{hyperref}
\usepackage{url}
\usepackage{comment}
\usepackage{pdfpages}
\usepackage{booktabs}
\usepackage{siunitx}

\usepackage{silence}
\bibpunct{(}{)}{;}{a}{}{,}             

\newcommand{\mum}{\textmu{}m\xspace}

\makeatletter
\renewcommand*\aa@pageof{, page \thepage{} of \pageref*{LastPage}}
\makeatother

\newcommandtwoopt{\citeads}[3][][]{\href{http://adsabs.harvard.edu/abs/#3}%
{\def\hyper@linkstart##1##2{}%
 \let\hyper@linkend\@empty\citealp[#1][#2]{#3}}}
 \newcommandtwoopt{\citepads}[3][][]{%
\href{http://adsabs.harvard.edu/abs/#3}{%
\def\hyper@linkstart##1##2{}%
\let\hyper@linkend\@empty~%
\citep[#1][#2]{#3}}%
}
\newcommandtwoopt{\citetads}[3][][]{\href{http://adsabs.harvard.edu/abs/#3}%
{\def\hyper@linkstart##1##2{}%
 \let\hyper@linkend\@empty\citet[#1][#2]{#3}}}
\newcommandtwoopt{\citeyearads}[3][][]%
{\href{http://adsabs.harvard.edu/abs/#3}
{\def\hyper@linkstart##1##2{}%
 \let\hyper@linkend\@empty\citeyear[#1][#2]{#3}}}

\newcommand{\HII}{\textrm{H~{\textsc{ii}}}\xspace}
\newcommand{\molh}{H$_2$\xspace}

\newcommand{\oric}{\mbox{$\theta^1$ Ori C}\xspace}
\newcommand{\Ta}{\HII\,region\xspace}
\newcommand{\Tb}{atomic PDR\xspace}
\newcommand{\Tc}{DF~1\xspace}
\newcommand{\Td}{DF~2\xspace}
\newcommand{\Te}{DF~3\xspace}

\newcommand{\usurfacebrightnessalt}{${\rm W\, m^{-2}\,sr^{-1}}$\xspace}

\begin{document}

\title{PDRs4All\\ {XIX.} The 6 to 9~\mum region as a probe of PAH charge and size in the Orion Bar}

\author{
    Baria Khan \inst{\ref{WOntarioPA}} \and 
    Samuel A. Daza Rodriguez \inst{\ref{Columbia}, \ref{WOntarioPA}} \and 
    Els Peeters\inst{\ref{WOntarioPA}, \ref{WOntarioIESE}, \ref{CarlSagan}} \and
    Alexander G.~G.~M. Tielens \inst{\ref{Leiden}, \ref{Maryland}} \and
    Takashi Onaka \inst{\ref{TokyoAstro}} \and
    Jan Cami \inst{\ref{WOntarioPA}, \ref{WOntarioIESE}, \ref{CarlSagan}} \and
    Bethany Schefter \inst{\ref{WOntarioPA}, \ref{WOntarioIESE}} \and
    Christiaan Boersma \inst{\ref{AMES}} \and
    Felipe Alarcón \inst{\ref{Milano}} \and
    Olivier Berné \inst{\ref{ToulouseIRAP}} \and
    Amélie Canin \inst{\ref{ToulouseIRAP}} \and
    Ryan Chown \inst{\ref{OhioState}, \ref{WOntarioPA}, \ref{WOntarioIESE}} \and
    Emmanuel Dartois \inst{\ref{SaclayISM}} \and
    Javier R. Goicoechea \inst{\ref{MadridFisFundamental}} \and
    Emilie Habart \inst{\ref{SaclayIAS}} \and
    Olga Kannavou \inst{\ref{SaclayIAS}} \and
    Alexandros Maragkoudakis \inst{\ref{AMES}} \and
    Amit Pathak \inst{\ref{BHU}} \and
    Alessandra Ricca \inst{\ref{AMES}, \ref{CarlSagan}} \and
    Gaël Rouillé \inst{\ref{Jena}} \and
    Dinalva A. Sales \inst{\ref{RioGrande}} \and
    Ilane Schroetter \inst{\ref{ToulouseIRAP}} \and
    Ameek Sidhu \inst{\ref{WOntarioPA}, \ref{WOntarioIESE}} \and
    Boris Trahin \inst{\ref{SaclayIAS}, \ref{STScI}} \and
    Dries Van De Putte \inst{\ref{WOntarioPA}, \ref{WOntarioIESE}, \ref{STScI}} \and
    Yong Zhang \inst{\ref{Sunyatsen}} \and
    Henning Zettergren \inst{\ref{Stockholm}}}
\institute{
    Department of Physics \& Astronomy, The University of Western Ontario, London ON N6A 3K7, Canada \label{WOntarioPA} \and 
    Universidad Nacional de Colombia, Ave Cra 30 $\#$ 45-3, Bogotá, Colombia \label{Columbia} \and 
    Institute for Earth and Space Exploration, The University of Western Ontario, London ON N6A 3K7, Canada
    \label{WOntarioIESE}          \and 
    Carl Sagan Center, SETI Institute, 339 Bernardo Avenue, Suite 200, Mountain View, CA 94043, USA
    \label{CarlSagan} \and 
  Leiden Observatory, Leiden University, P.O. Box 9513, 2300 RA Leiden, The Netherlands
  \label{Leiden} \and
  Astronomy Department, University of Maryland, College Park, MD 20742, USA
  \label{Maryland} \and
  Department of Astronomy, Graduate School of Science, The University of Tokyo, 7-3-1 Bunkyo-ku, Tokyo 113-0033, Japan
  \label{TokyoAstro}\and
  NASA Ames Research Center, MS 245-6, Moffett Field, CA 94035-1000, USA
  \label{AMES} \and  
   Dipartimento di Fisica, Università degli Studi di Milano, Via Celoria 16, 20133 Milano, Italy
   \label{Milano} \and
     Institut de Recherche en Astrophysique et Planétologie, Université Toulouse III - Paul Sabatier, CNRS, CNES, 9 Av. du colonel Roche, 31028 Toulouse Cedex 04, France
  \label{ToulouseIRAP} \and
  Astronomy Department, Ohio State University, Columbus, OH 43210 USA
  \label{OhioState} \and
  Institut des Sciences Moléculaires d'Orsay, Université Paris-Saclay, CNRS, Bâtiment 520, 91405 Orsay Cedex, France
  \label{SaclayISM} \and
  Instituto de Física Fundamental (CSIC), Calle Serrano 121-123, 28006, Madrid, Spain
  \label{MadridFisFundamental} \and
  Institut d'Astrophysique Spatiale, Université Paris-Saclay, CNRS,  Bâtiment 121, 91405 Orsay Cedex, France
  \label{SaclayIAS} \and
  Department of Physics, Institute of Science, Banaras Hindu University (BHU), Varanasi, 221005, India  \label{BHU}
\and 
   Astrophysical Institute and University Observatory, Schillergässchen 2-3, 07745 Jena, Germany \label{Jena}  \and
  Instituto de Matemática, Estatística e Física, Universidade Federal do Rio Grande, 96201-900, Rio Grande, RS, Brazil
  \label{RioGrande}   \and
  Space Telescope Science Institute, 3700 San Martin Drive, Baltimore, MD 21218, USA \label{STScI} \and
  School of Physics and Astronomy, Sun Yat-sen University, 2 Da Xue Road, Tangjia, Zhuhai 519000,  Guangdong Province, China
  \label{Sunyatsen}\and
   Atomic Physics Division, Stockholm University, SE-106 91 Stockholm, Sweden \label{Stockholm} }

\titlerunning{PAH charge distributions in the Orion Bar}
\authorrunning{Khan et al.}

   \date{Received xxx; accepted xxx}

  \abstract
  {Infrared emission from polycyclic aromatic hydrocarbons (PAHs) play a major role in determining the charge balance of their host environments that include photo-dissociation regions (PDRs) in galaxies, planetary nebulae, and rims of molecular clouds.}
  {We aim to investigate the distribution and sizes of charged PAHs across the key zones of the Orion Bar PDR; i.e., the ionization front, the \Tb and the dissociation fronts.}  %
  {We employ JWST MIRI-MRS observations of the Orion Bar from the Early Release Science program "PDRs4All" and synthetic images in the JWST MIRI filters. We investigate the spatial morphology of the AIBs at 6.2, 7.7, 8.6, and 11.0~\mum that commonly trace PAH cations, and the neutral PAH-tracing 11.2~\mum AIB, their (relative) correlations, and the relationship with existing empirical prescriptions for AIBs.}
   {The 6.2. 7.7, 8.6, 11.0, and 11.2~\mum AIBs are similar in spatial morphology, on larger scales. Aside from the 11.0~\mum AIB, these AIBs exhibit enhanced intensities at the dissociation fronts. Analyzing three-feature intensity correlations, two distinct groups emerge: the 8.6 and 11.0~\mum vs. the 6.2 and 7.7~\mum AIBs. 
   We attribute these correlations to PAH size. The 6.2 and 7.7~\mum AIBs trace cationic, medium-sized PAHs. Quantum chemical calculations reveal that the 8.6~\mum AIB is carried by large, compact, cationic PAHs, and the 11.0~\mum AIB's correlation to it implies, so is this band. The 6.2/8.6 and 7.7/8.6 PAH band ratios thus probe PAH size. We conclude that the 6.2/11.2 AIB ratio is the most reliable proxy for charged PAHs, within the cohort. We outline JWST MIRI imaging prescriptions that serve as effective tracers of the PAH ionization fraction as traced by the 7.7/11.2 PAH emission.}
   {This study showcases the efficacy of the 6-9~\mum AIBs to probe the charge state and size distribution of the emitting PAHs, offering insights into the physical conditions of their host environments. JWST MIRI photometry offers a viable alternative to IFU spectroscopy for characterizing this emission in extended objects.}
\maketitle

\section{Introduction}

Polycyclic aromatic hydrocarbons (PAHs) comprise a prominent and influential family of organic molecules in space. PAHs are the dominant carriers of the strong, broad aromatic infrared bands (AIBs), the most prominent of which occur at 3.3, 6.2, 7.7, 8.6, 11.2, and 12.7~\mum , that have been detected across the Universe~\citep[e.g.,][]{Gillett1973, Merrill1975, Allamandola1985, Peeters2002}. A remarkably diverse set of astrophysical environments have been found to exhibit AIBs, including the photodissociation regions (PDR) environments of the interstellar medium, planetary nebulae, reflection nebulae and star-forming galaxies~\citep{Tielens2008, Li2020}. By virtue of their prevalence and molecular nature, PAHs play a crucial
role in shaping the physical and chemical landscapes of the interstellar medium, and have important
diagnostic value for understanding the physical conditions of their host environments~\citep[e.g.,][]{Galliano2008, peeters2017, OBpahs, Schroetter2024}.

\begin{figure*}
\begin{center}
\resizebox{.99\hsize}{!}{
\includegraphics{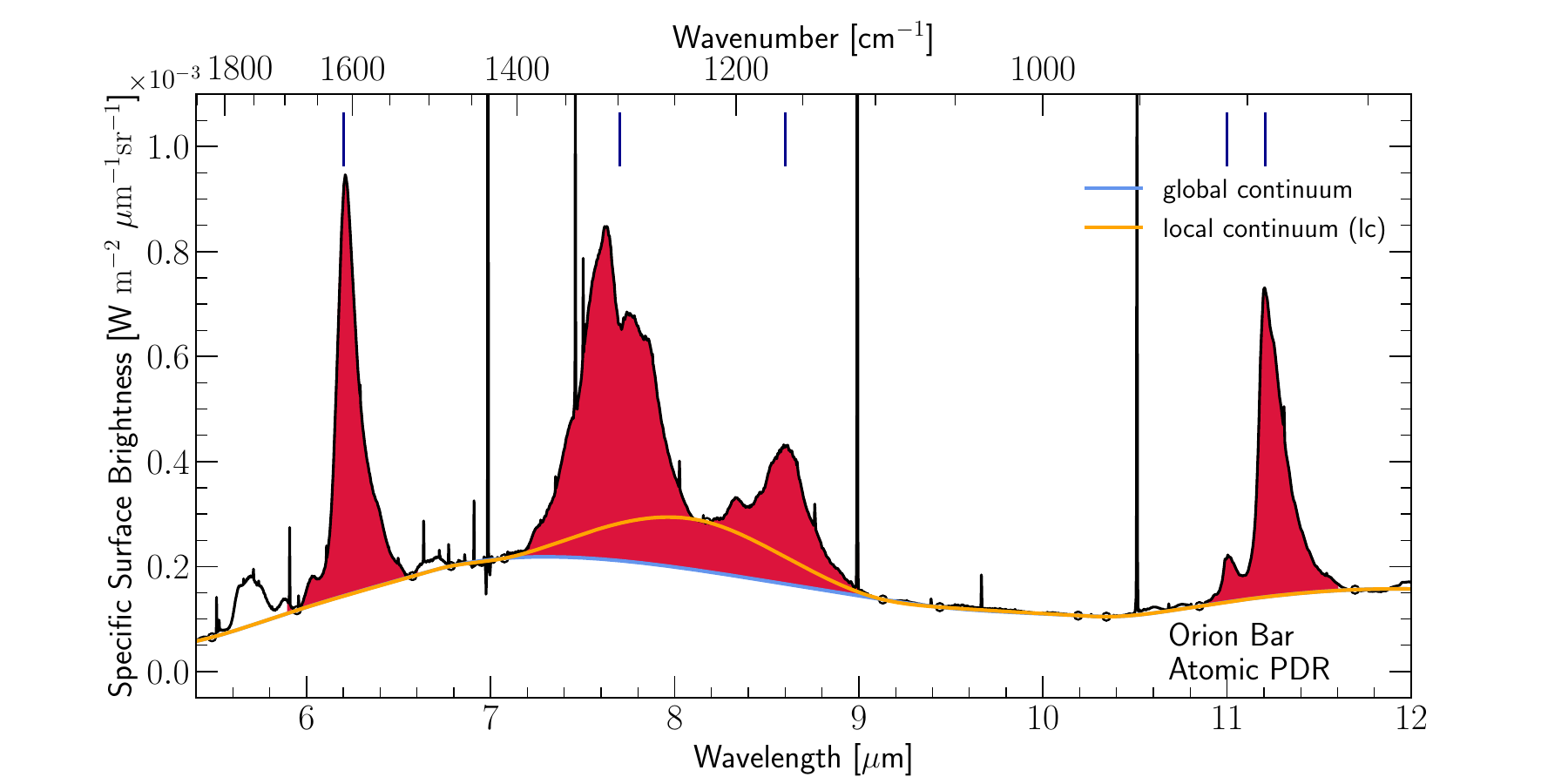}
}
\caption{The AIB template spectrum for the Orion Bar Atomic PDR, showcasing the 6.2, 7.7, 8.6, 11.0, and 11.2~\mum AIBs shaded in red and indicated by the vertical solid blue lines. The blue curve shows the underlying  "global" continuum, the orange curve is the "local" continuum (lc), and the black round symbols mark the anchor points used for the continua determination.}
\label{fig:ADPR_spec}

\end{center}
\end{figure*}

While PAH emission is remarkably constant across diverse environments,
nuanced variations in relative band strengths and profiles in the observed
spectra contain valuable information about distinct molecular characteristics
such as size, charge, and molecular edge structure, of the astronomical PAH population~\citep[e.g.,][]{hony2001, Peeters2002, Galliano2008, shannon2016}. In particular, the charge state of PAHs is a main driver of variations in the spectra of PAHs~\citep{Allamandola1999, Oomens2003, ricca2012,peeters2017, sidhu2021}. The 6-9~\mum C-C stretching and C-H in-plane bending modes are strongly enhanced in ionized PAHs, resulting in the well-established strong correlations of prominent 6.2, 7.7, and 8.6~\mum AIBs. In contrast, the neutral PAHs display strong C-H modes in the 3.3 and 11-15~\mum regions, which includes the nominal 11.2~\mum AIB ascribed to neutral, solo edge hydrogen dominated PAHs~\citep{Allamandola1999, Candian2014}. 
This dichotomy in PAH band strengths thus provides a diagnostic of the ionization state of the PAH population, and thereby of the local physical conditions through the ionization parameter, $\gamma$~$=$ $G_0$/$n_\mathrm{e}T_\mathrm{gas}^{1/2}$, where $G_0$ is the strength of the local UV radiation field in units of the Habing field~\citep{Habing1968}, $n_\mathrm{e}$ is the electron density, and $T_\mathrm{gas}$ is the gas temperature~\citep{Bregman2005,Galliano2008}. These conclusions were based upon {ISO/SWS}\footnote{The Short-Wavelength Spectrometer (SWS) on-board the European Space Agency (ESA)'s Infrared Space Observatory (ISO).} and {Spitzer/IRS}\footnote{The Infrared Spectrograph (IRS) on the Spitzer Space Telescope.} observations, which had limited spatial and spectral resolution. The high spatial and spectral resolution observations now available with the James Webb Space Telescope (JWST) MIRI Medium Resolution Spectroscopy (MRS) Integral Field Unit (IFU) instrument, allow us to reassess the viability of these PAH charge proxies. We also aim to address the viability of using JWST broad band filter observations to address PAH properties in extragalactic observations. To accomplish this, we study the traditional PAH charge proxy AIBs at 6.2, 7.7, 8.6, 11.0~\mum and the canonically neutral 11.2~\mum AIB, as observed by the JWST in the Orion Bar, a well-studied PDR.

We present the details of the observations in Sect.~\ref{Observations}. The spatial behavior and correlations of the AIBs in our cohort are presented in Sect.~\ref{Results}. We discuss the efficacy of the 7--9~\mum AIBs as size and charge proxies in Sect.~\ref{discussion_01} and Sect.~\ref{discussion_02}, respectively. We discuss the use of JWST photometric observations to probe the ionic PAH population in Sect.~\ref{discussion_03}. Finally, a summary of the results of this study is given in Sect.~\ref{sec:conclusion}.

\section{Observations \& Analysis}
\label{Observations}

Within the Orion Nebula, the massive star-forming region closest to us, lies the Orion Bar PDR, wherein far-ultraviolet (FUV) radiation from the brightest member of the Trapezium cluster of massive stars, \oric, drives the physical and chemical conditions of the neutral gas~\citep{Tielens1985b}. The Orion Bar PDR straddles the boundary between the surface of the \HII region and the surrounding molecular cloud~\citep{Bally2008}. The PDR is viewed nearly edge-on and its physical and chemical stratification has now been seen in spectacular detail through JWST NIRCam, NIRSpec, and MIRI imaging
and spectroscopic observations~\citep{OBim, OBspec, OBpahs, OBlines}. We present a NIRCam image of the Orion Bar in Fig.~\ref{fig:FOV}. The PDR begins just beyond the sharp ionization front (IF), where the gas becomes neutral and predominantly atomic. As the FUV photon flux attenuates with distance into the PDR, the gas becomes molecular. This marks the beginning of the molecular PDR. Herein, the spectral emission from \molh exhibits several ridges at increasing distances from the IF. In the MIRI field of view (FOV), three dissociation fronts (DFs) (\Tc, \Td and \Te) are observed. We refer the interested reader to \citet{OBim}, \citet{OBspec}, and \citet{OBhydrocarbon} for a detailed description of the geometry and large-scale stratification of the Orion Bar PDR. 
The edge-on geometry and close proximity make the Orion Bar an ideal target for PDR studies, investigating dust and gas photoprocessing with respect to distance from the ionizing source and the photochemical evolution of the resident PAHs.

We utilize the JWST MIRI-MRS IFU~\citep{MIRI} observations of the Orion Bar made for the JWST Early Release Science (ERS) program
"PDRs4All: Radiative Feedback from Massive Stars"~\citep{PDRs4AllPASP}. The MIRI-MRS data spans the 4.90–27.90~\mum wavelength range at a resolving power of~$\sim$1500--3500. The details of data reduction and production of the spatio-spectral mosaic used in this study are provided in~\cite{OBlines} and~\cite{OBpahs}.

\begin{figure*}[h!]
    \begin{center}
        \resizebox{1.\hsize}{!}{
    \includegraphics{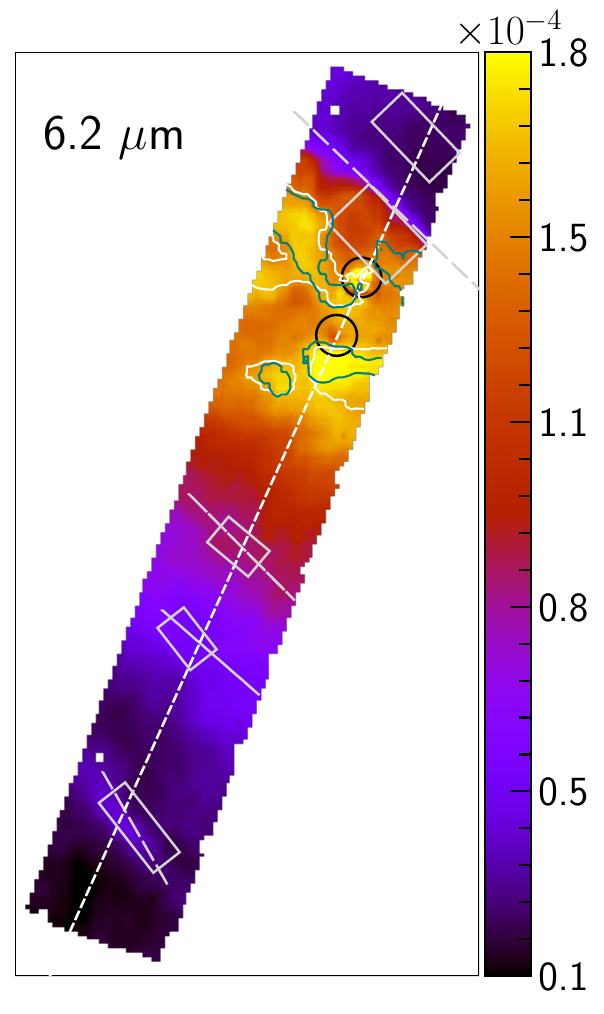}
    \includegraphics{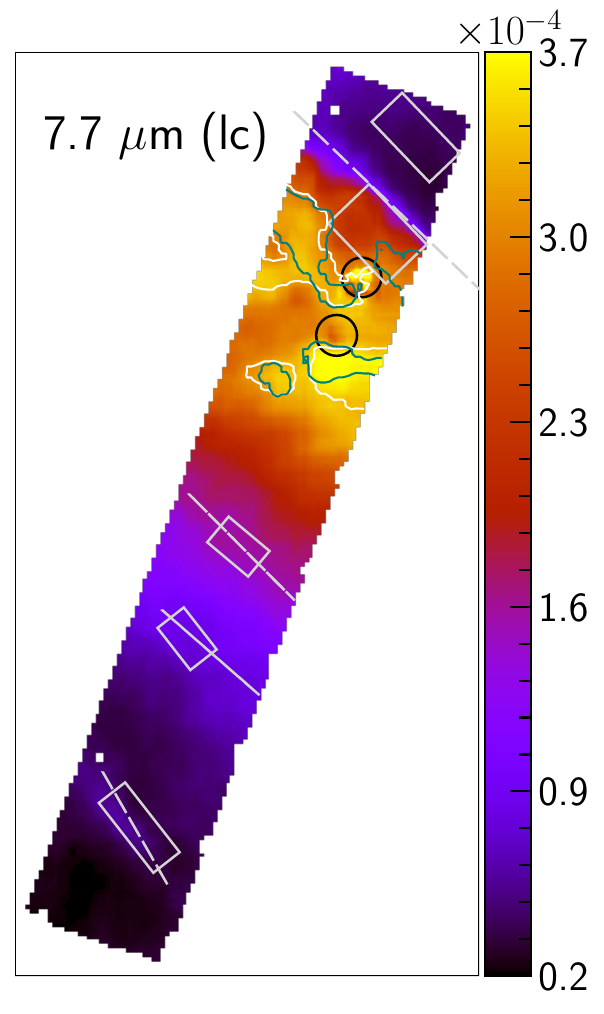}
    \includegraphics{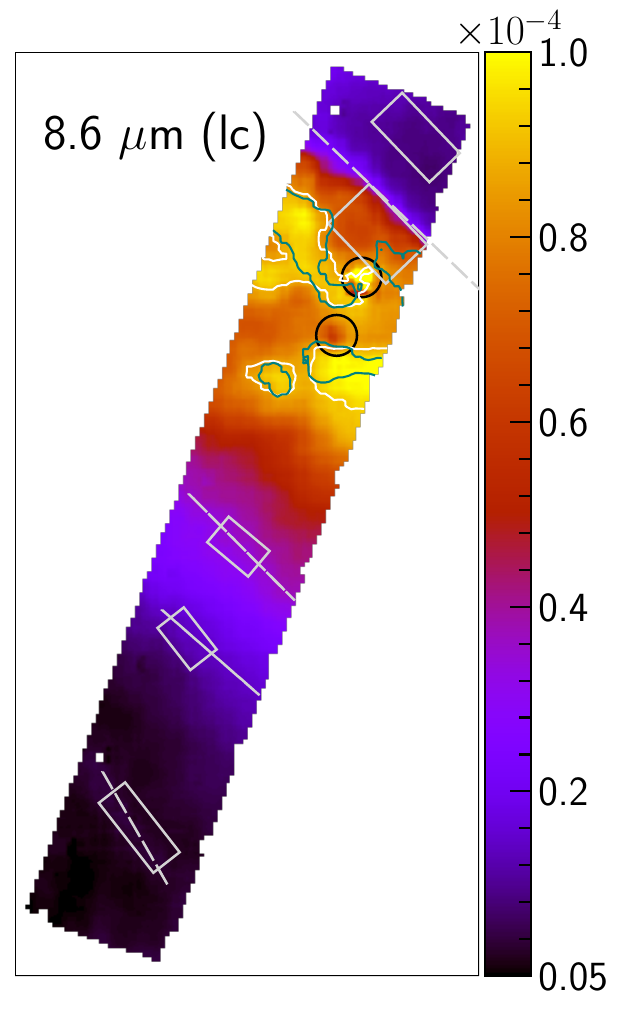}
    \includegraphics{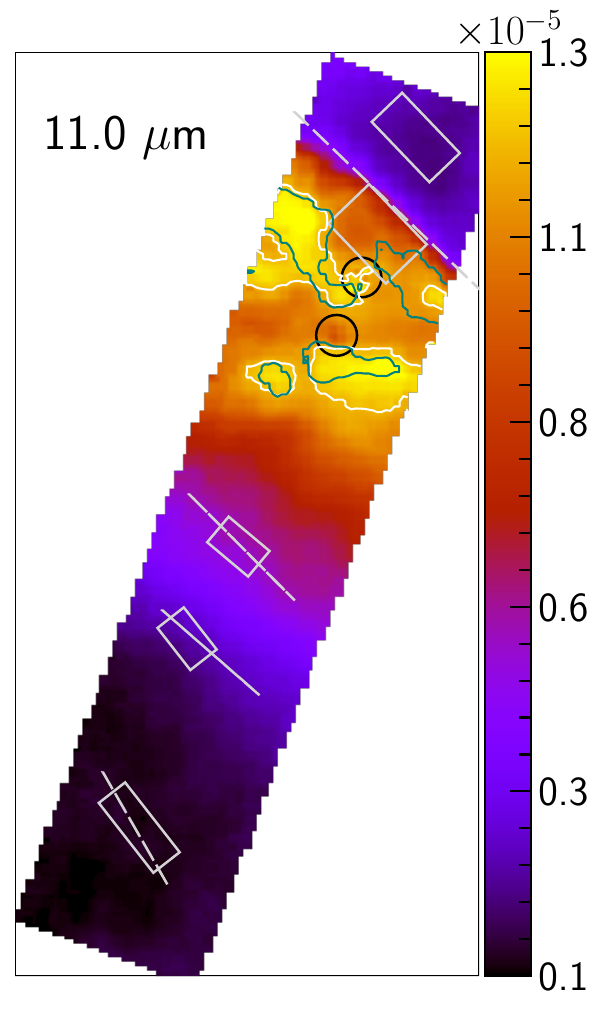}
    \includegraphics{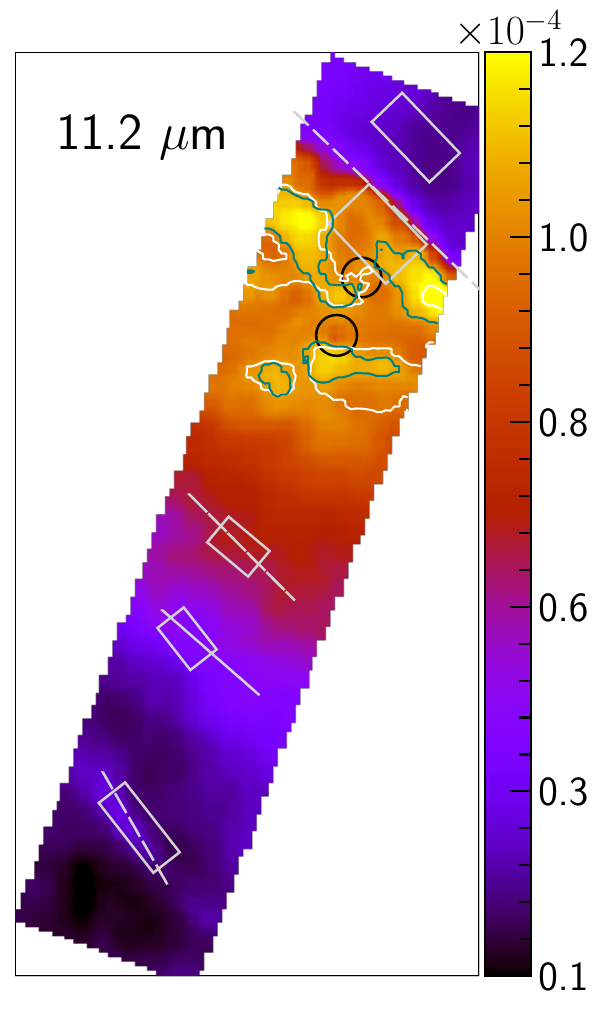}}
        \resizebox{.8\hsize}{!}{
    \includegraphics{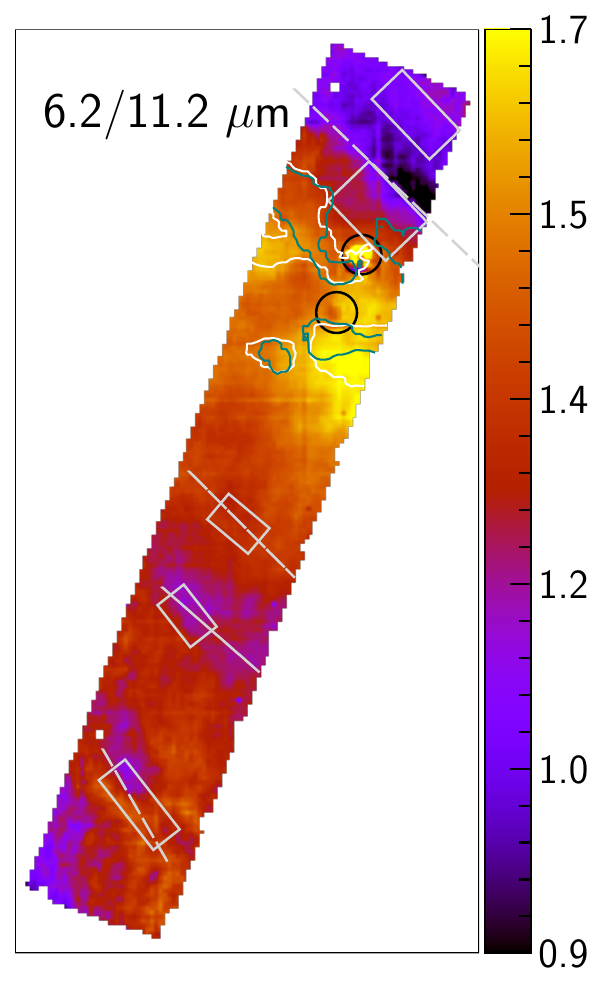} 
    \includegraphics{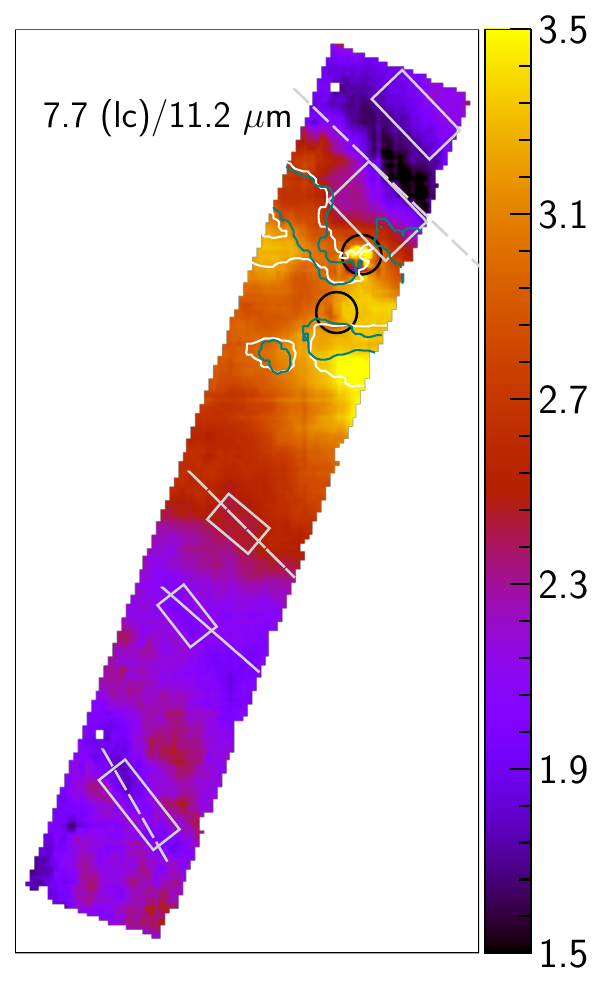}    
    \includegraphics{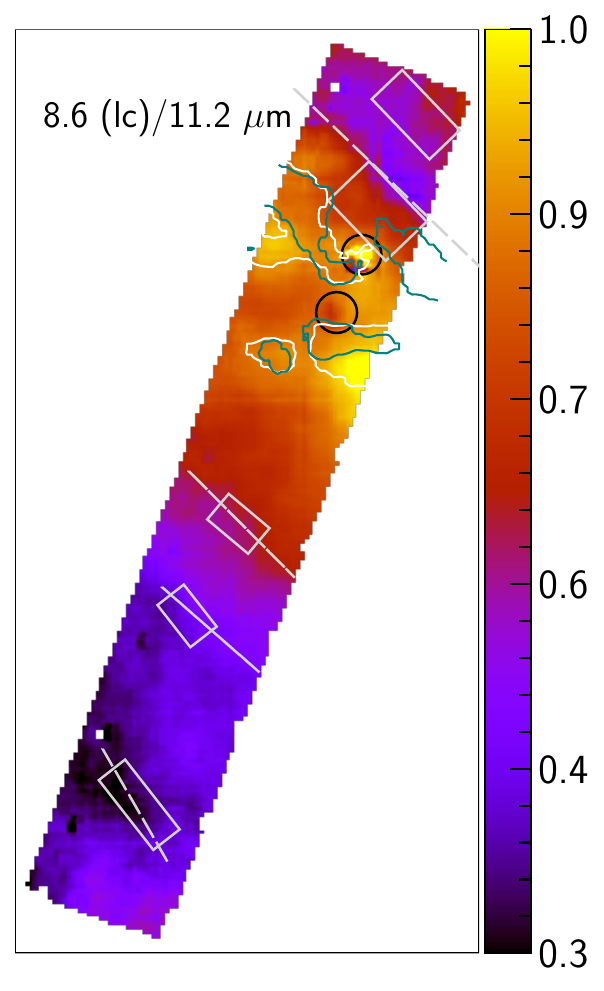}
    \includegraphics{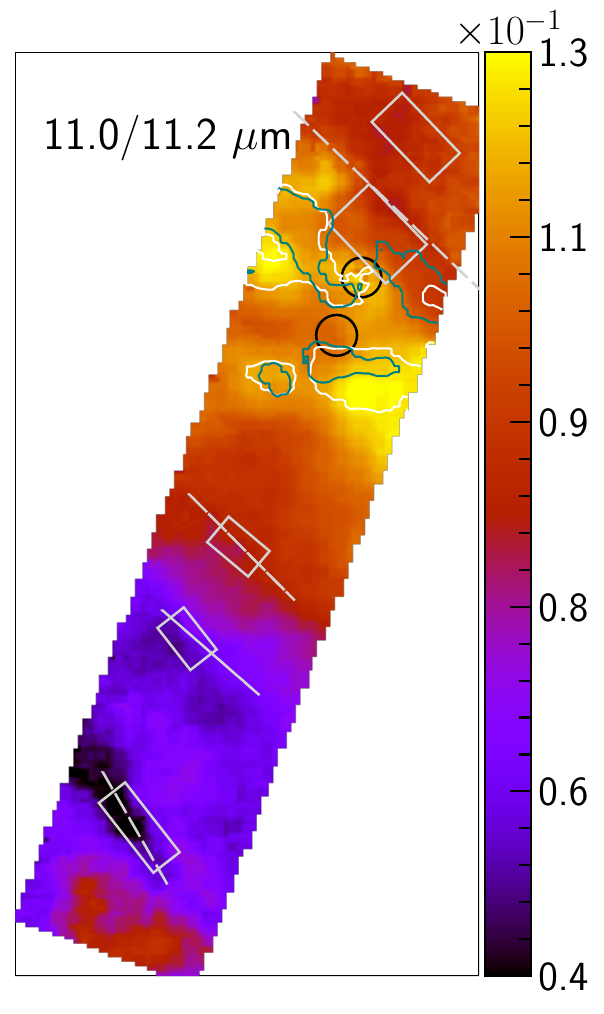}
    }
    \caption{Spatial distribution of the surface brightnesses of the 6.2, 7.7, 8.6, 11.0, and 11.2~\mum AIBs in the Orion Bar PDR, in units of \usurfacebrightnessalt, and brightness ratios relative to the 11.2~\mum AIB. The analysis utilizes the 7.7 and 8.6~\mum  bands measured using a local continuum (lc; Fig.~\ref{fig:ADPR_spec}).
    \oric is located toward the top right of each map. For each map, the range of the corresponding color bar is set between 0.5$\%$ and 99.5$\%$ percentile level for the data, while zero pixels, edge pixels, and pixels covering the two proplyds as seen in the MIRI mosaic, indicated by the black circles, are masked out. The contours trace peak emission
    for the 11.0~\mum AIB (white) and the 11.2~\mum AIB (teal). The rectangular apertures of the template spectra for the \Ta, the \Tb, \Tc, \Td, and \Te, from top to bottom,
    are shown in gray, the gray lines delineate the IF and the three dissociation fronts, \Tc, \Td and \Te, and the dashed, diagonal white line in the top left map indicates the cut across the MIRI mosaic (position angle of 155.79°).}
    \label{fig:maps}
    \end{center}
\end{figure*}

\begin{figure}[h!]
    \begin{center}
   \resizebox{1\columnwidth}{!}{
   \includegraphics[]{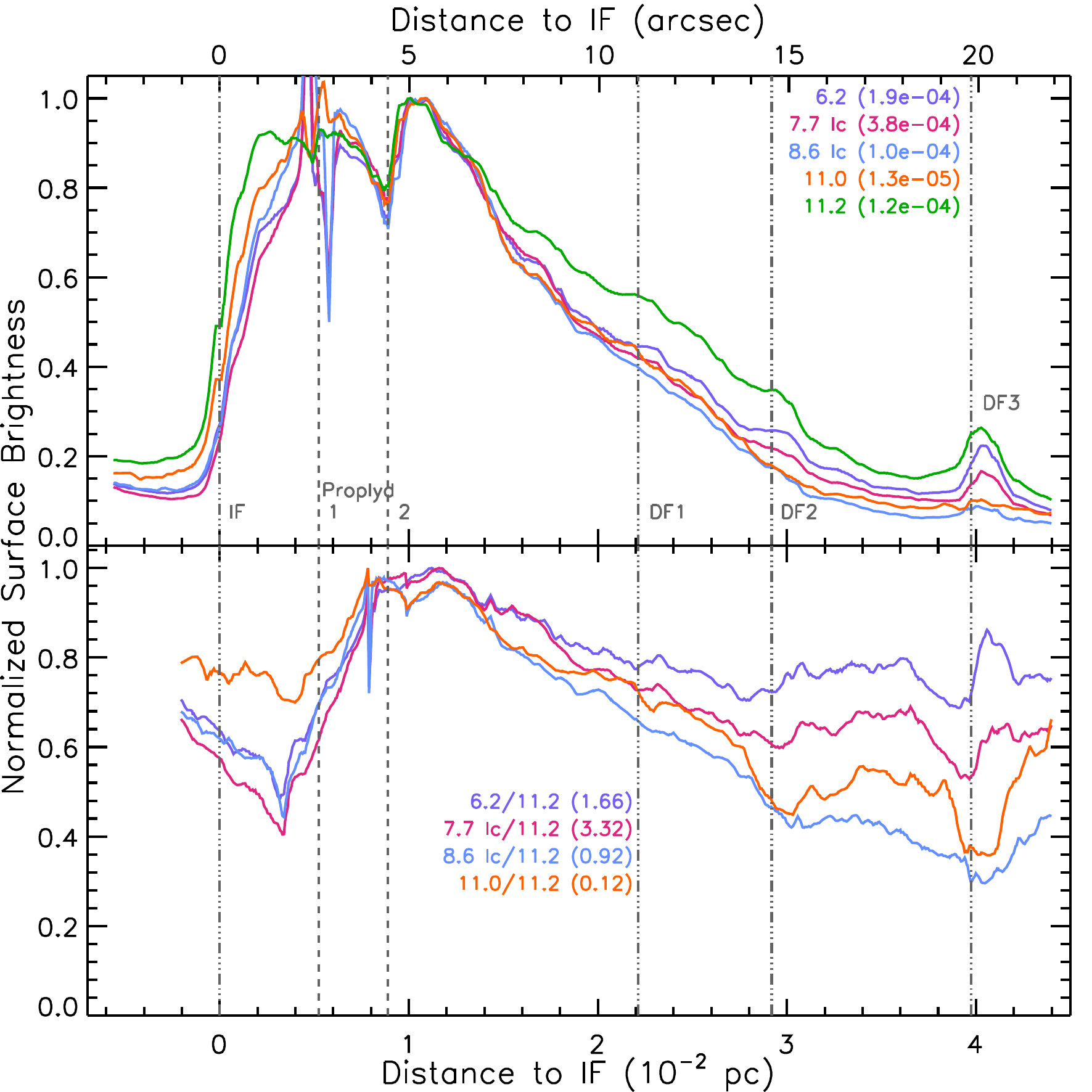}}
   \caption{Normalized surface brightnesses and their ratios for the 6.2, 7.7, 8.6, 11.0, and 11.2~\mum AIBs as a function of distance to the IF (0.228 pc or 113.4\arcsec\, from \oric) along a cut crossing the mosaic (see Fig.~\ref{fig:maps} for the location of the cut). Normalization factors are listed in \usurfacebrightnessalt in parentheses for each surface brightness. As the cut is not perpendicular to the IF and distances are given along the cut, a correction factor of cos(19.58$\degr$)~$=$ 0.942 needs to be applied to obtain a perpendicular distance from the IF.}
   \label{fig:cuts}
    \end{center}
\end{figure}

To investigate variations in the spatial distributions of the charge-tracing AIBs (at 6.2, 7.7, 8.6, and 11.0~\mum) and study their correlations, the specific surface brightnesses of the AIBs were integrated for the continuum-subtracted spectra in the mosaic. For this purpose, as shown in Fig.~\ref{fig:ADPR_spec}, we consider a "local" continuum (lc; orange), for which an anchor point is placed at roughly 8.2~\mum, as has been done previously by~\cite{hony2001} and~\cite{Peeters2002}, for example. 
We also estimate a "global" spline continuum (blue) upon which the AIBs lie superposed. The wavelength integration windows are provided in Appendix~\ref{asubsec:ranges}.  We direct the interested reader to~\cite{OBOOPs} for details on the methodology of the study.

\section{Results}
\label{Results}

We present the spatial distribution and surface brightnesses of the 6.2, 7.7, 8.6, 11.0, and 11.2~\mum AIBs, and distributions relative to the 11.2~\mum AIB, in Fig.~\ref{fig:maps}. The radial profiles of the surface brightnesses of the AIBs, measured along a cut across to the Orion Bar (Fig.~\ref{fig:maps}), are presented in Fig.~\ref{fig:cuts}. In the following discussion, we refer to results based on the local continuum unless otherwise specified. The results using the global continuum are presented in Appendix~\ref{asubsec:more_maps}. The spatial behavior of the AIBs in this cohort is akin to the 3.2--3.7~\mum and 10--15~\mum AIBs investigated by~\cite{OBspec} and~\cite{OBOOPs}. The AIB 
emission observed in the \Ta arises from the background face-on PDR formed at the surface of the molecular cloud OMC-1. The strongest AIB emission is observed within the \Tb that marks the first, predominant layer of the PDR, and begins just beyond the IF demarcating the regions of ionized and neutral gas. The AIB fluorescence is widespread throughout the \Tb, with a few regions of local maximum emission. As the FUV photon flux attenuates with distance from the IF, the AIB emission attenuates similarly. However, local enhancements in the AIB strengths are observed near the DFs in the molecular zone of the Orion Bar.
At large, the 6.2, 7.7, 8.6, 11.0, and 11.2~\mum maps are all very similar. As seen in Fig.~\ref{fig:cuts}, on smaller scales, as for the 11.2~\mum and 3.3~\mum AIB \citep[see][]{OBspec}, the 6.2, 7.7, and 8.6~\mum AIB ensemble emission is enhanced just beyond each of the \molh DFs. The 11.0~\mum emission experiences no such local peaks near the DFs.

Historically, the PAH ionization fraction has been mapped through the 6.2/11.2, 7.7/11.2, and 8.6/11.2 surface brightness ratios. The maps in the lower panels of Fig.~\ref{fig:maps} reveal that the PAH ionization fraction is the largest on the north-western end of the \Tb. Figure~\ref{fig:cuts} reveals that the 6.2/11.2, 7.7/11.2, and 8.6/11.2 relative surface brightnesses are generally larger than the 11.0/11.2 ratio just past the proplyds in the \Tb. The normalized surface brightness ratio of 6.2/11.2 shows a lesser decrease with distance from the IF than the 7.7/11.2 and 8.6/11.2 ratios. The 6.2/11.2 and 7.7/11.2 ratios 
get enhanced just past each of the three DFs, unlike the 8.6/11.2 and 11.0/11.2 surface brightness ratios that drop at these fronts.

\begin{figure*}[ht!]
    \begin{center}
    \begin{tabular}{c}
        \vspace{-.5em}        \includegraphics[width=0.25\textwidth]{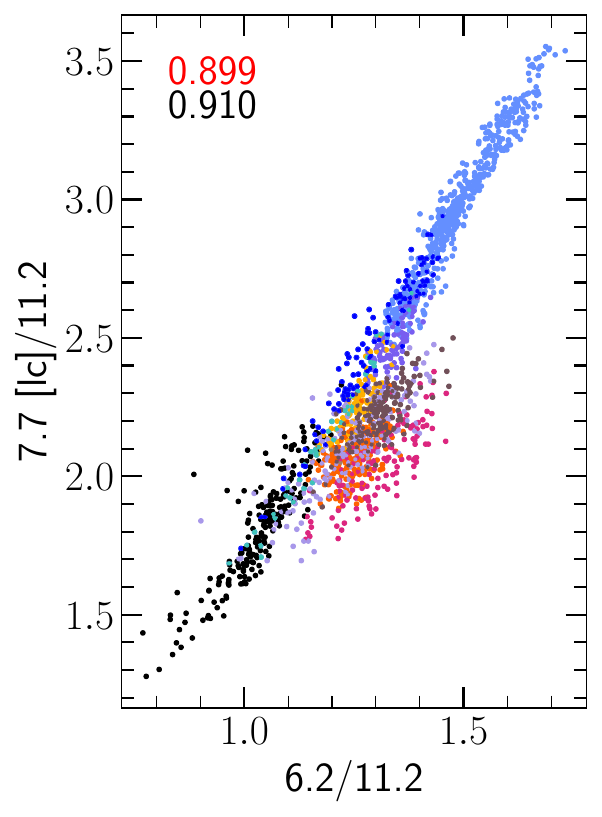}
            \includegraphics[width=0.25\textwidth]{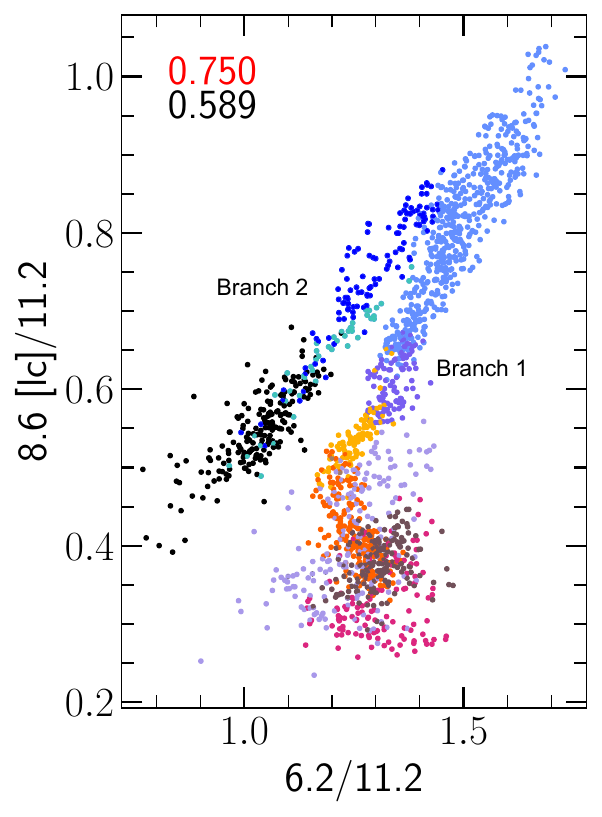}
            \includegraphics[width=0.25\textwidth]{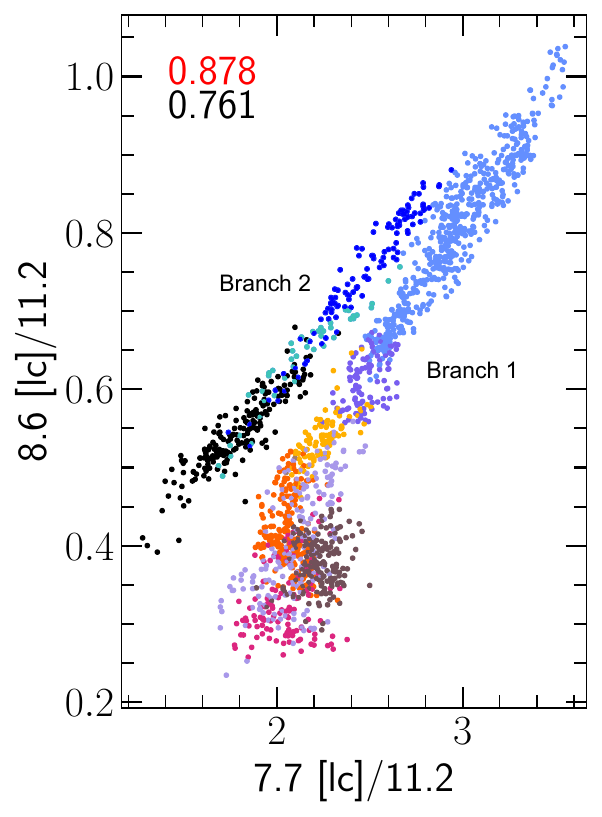}
            \raisebox{.8cm}[0pt][0pt]
            {\includegraphics[height=0.22\textheight]{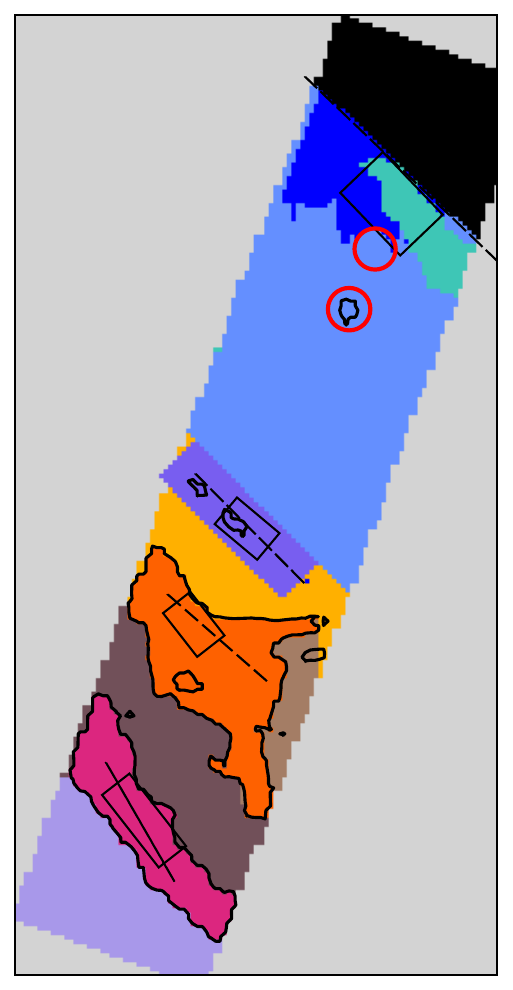}}
     \vspace{-.2em} \\

            \includegraphics[width=0.25\textwidth]{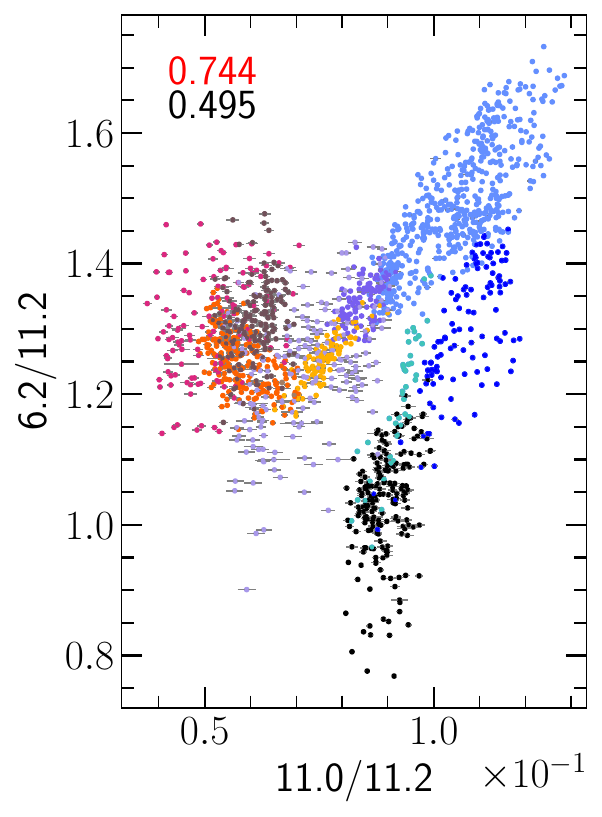}
             
            \includegraphics[width=0.25\textwidth]{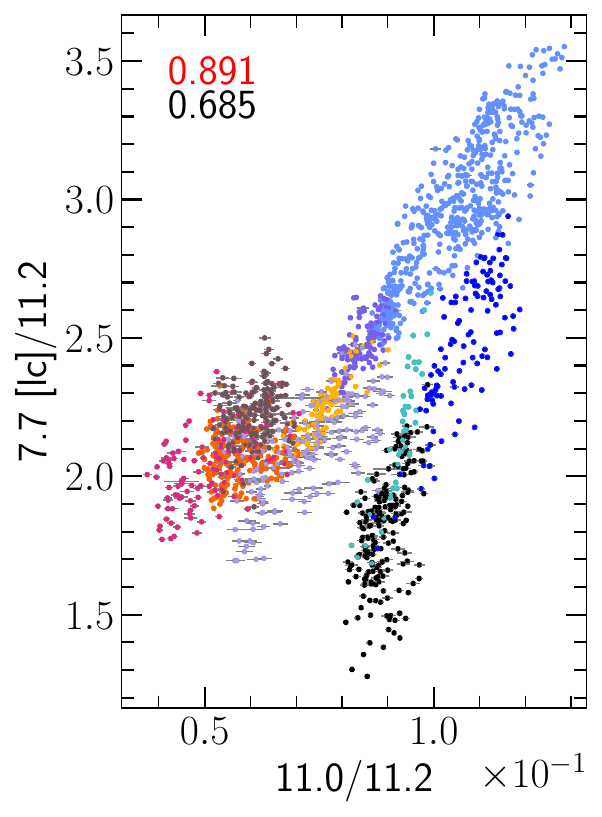}
            \includegraphics[width=0.25\textwidth]{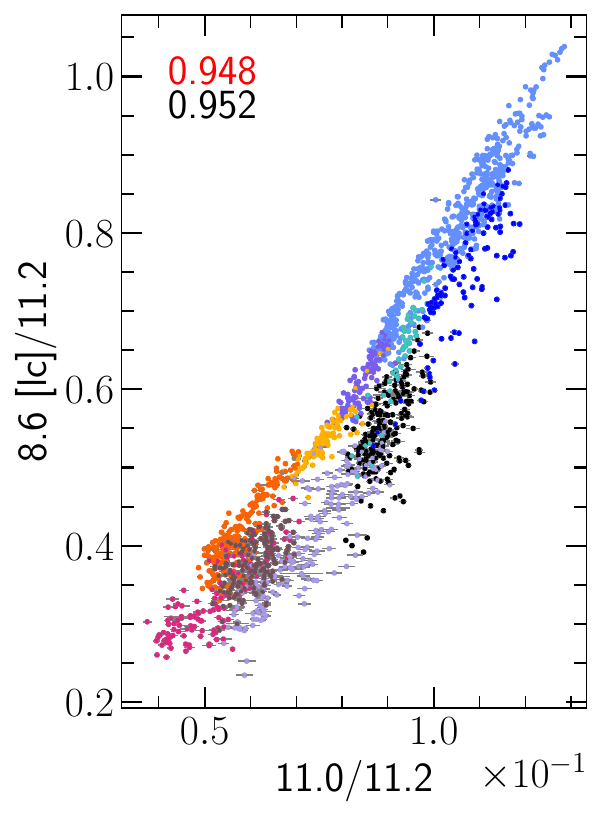}
       
    \end{tabular}
 \caption{Correlations of the 7.7 [lc]/11.2, 8.6 [lc]/11.2, and 11.0/11.2 AIB surface brightness ratios. The Spearman correlation coefficients `R’ for the variables excluding
(including) the points from the \HII region (black), and regions beneath the IF (teal and dark blue), are printed in red (black) in the panels. Only surface brightnesses from every other spaxel are considered in the correlation analyses. The data points are colored according to regions in the mosaic where those pixels are located (top right panel).  This visual region-color scheme is as follows: blue (atomic PDR), dark blue (east IF), teal (west IF), purple (\Tc), yellow (region between \Tc and \Td), orange (\Td), light brown (region west of DF 2), brown (region between \Td and \Te), pink (\Te), and light purple (region beyond DF 3). The contours highlighting \Td and \Te are identified through enhanced emission of \molh 0-0 S(3) in these DFs. The red circular regions on the color-region map (top right panel) correspond to the protoplanetary disks within the line-of-sight of observations, which have been masked in this analysis.}
    \label{fig:corr}
    \end{center}
\end{figure*}

We further examine the relationships between these AIBs through three-feature intensity correlation plots, where the AIB surface brightnesses are normalized to the 11.2~\mum AIB (Fig.~\ref{fig:corr}). Out of the 6.2, 7.7, and 8.6~\mum bands, the normalized 6.2 and 7.7~\mum AIB surface brightnesses exhibit the strongest correlation. 
The 8.6/11.2 vs. 7.7/11.2 and 8.6/11.2 vs. 6.2/11.2 correlations reveal branching. The main branch (`Branch 1') traces emission from the edge-on PDR that is, the majority of the \Tb and the DFs, while `Branch 2' comprises emission from the \Ta, regions just below the IF, and some contribution from the \Tb. We note that the 6.2 and 7.7~\mum AIB surface brightnesses are tightly correlated over all the regions, such that the branches coincide. Focusing on the edge-on PDR (the main branch), we observe a distinct change in slope, where the 8.6/11.2 ratio decreases between \Td and \Te, at \Te, and beyond \Te. This suggests that the 8.6~\mum AIB surface brightness in these regions is "missing" contributions from PAHs that do contribute to the 6.2 and 7.7~\mum AIBs. We conclude therefore that only a subset of the interstellar PAH family contributes to the 8.6~\mum AIB and that this subset has a relatively low abundance in the regions beyond \Td. We propose that this is linked to the intrinsic vibrational properties of PAHs, discussed in Sect.~\ref{discussion_01}. 
In addition, \cite{OBOOPs} suggested that the AIB emission observed in \Ta and near the IF — right at the precipice of the PDR — is contaminated by emission from the background face-on PDR and not from the edge-on PDR. The morphology of the Orion Bar thus drives the branching behavior seen in the correlation plots (Fig.~\ref{fig:corr}).

As to the 11.0~\mum band, Fig.~\ref{fig:corr} shows that it correlates best in brightness with the 8.6~\mum AIB, as the emission throughout the PDR until the edge of \Te follows a linear relationship. Comparatively, the correlations of the 7.7 and 6.2~\mum bands with the 11.0~\mum AIB are poorer, and display distinct branches.

\begin{figure}[ht!]
    \begin{center}
    \begin{tabular}{c}
        \vspace{-.5em}
        \resizebox{\hsize}{!}{
        \includegraphics[width=0.28\textwidth]{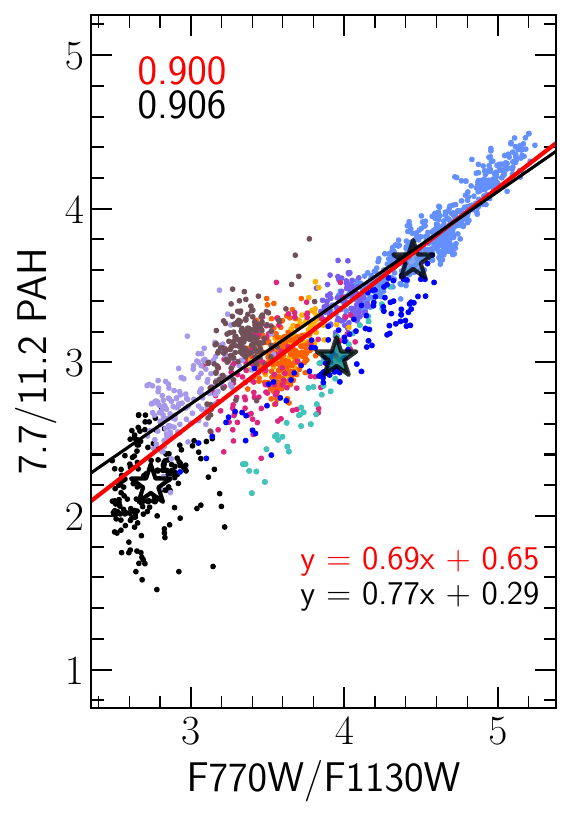}
            \includegraphics[width=0.27\textwidth]{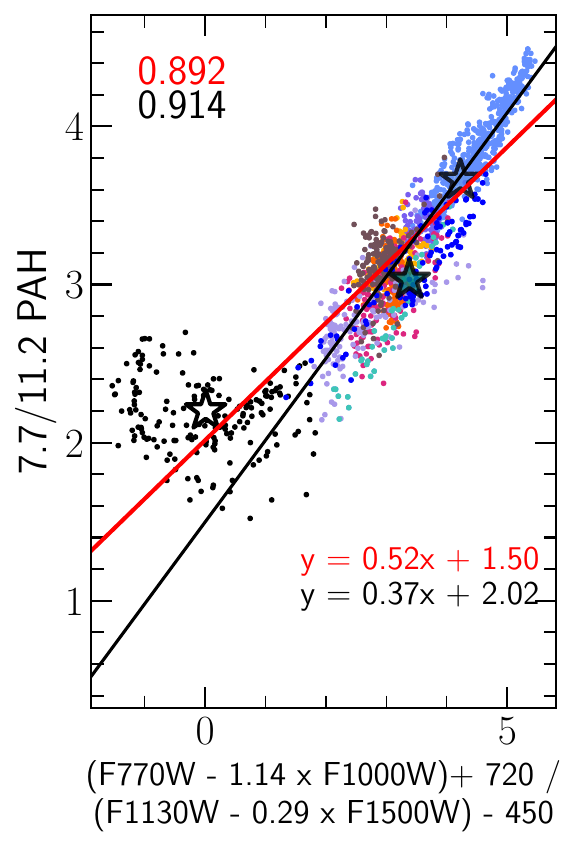}}
    \end{tabular}
    \end{center}
    \caption{Two strongest correlations between ratios of the synthetic images involving F770W, F1130W, and F1500W and the measured 7.7/11.2 spectroscopic PAH emission in the Orion Bar. Here the 7.7~\mum feature is measured using the global continuum (Fig.~\ref{fig:ADPR_spec}). Lines of best-fit through the data excluding
(including) the points from the \Ta (black), and the regions near the IF (teal and dark blue), their equations and the Spearman correlation coefficients R are shown in red (black). The star symbols highlight the average values for the \Tb (blue), the \Ta (black) and the region near the IF (teal and dark blue).}
    \label{fig:photom_spec}
\end{figure}

\section{Discussion}

\subsection{The 6.2/8.6 and 7.7/8.6 AIB ratios as tracers for PAH size}
\label{discussion_01}

Quantum chemistry studies have revealed that the 8.6~\mum C-H in-plane bending mode is typically not very distinct in the calculated emission spectra of irregular PAHs~\citep{Bauschlicher:2008, Bauschlicher2009}. This band only becomes prominent for large, compact regular PAH cations and anions and its intrinsic strength relative to the C-C modes at 6.2~\mum and 7.7~\mum increases with size~\citep{Bauschlicher:2008, Bauschlicher2009, ricca2012}. This behavior has been taken to imply that the 8.6~\mum AIB originates in large ($\approx$66-130 C-atoms), compact PAHs - such as the coronene family, characterized by only solo and duo edge hydrogens~\citep{Bauschlicher:2008, Pathak2008}. On the other hand, the 6.2 and 7.7~\mum AIBs are primarily associated with medium ($\approx$30-66 C-atoms) and large sized PAHs of any structure~\citep{Schutte1993, ricca2012, peeters2017}.
We emphasize then that the 6.2/8.6 and 7.7/8.6~\mum AIB ratios provide a size estimator that is independent of the well-known 3.3/11.2 size estimator~\citep{Allamandola1989, Croiset2016, Maragkoudakis2020}. However, in contrast to the latter, the 6.2/8.6 and 7.7/8.6~\mum ratios may only measure the size of the subset of compact PAHs in the interstellar PAH family. 

Large, symmetric PAHs, which are responsible for the strong 8.6~\mum AIB emission, are also inherently enriched in solo edge hydrogens~\citep{Pathak2008}. This naturally
connects the two cationic AIBs at 8.6~\mum and 11.0~\mum, where the latter similarly traces large PAHs, albeit through the C-H out-of-plane bending vibrations of solo edge hydrogens. The shared origin of the 8.6 and 11.0~\mum AIBs in large, compact PAHs is also supported by the observation of bright and co-spatial 8.6 and 11.0~\mum AIB emissions in close vicinity of the illuminating star of the reflection nebula NGC~2023. This results from the survival of the emitting large, resilient PAHs in the star's strong radiation field~\citep{peeters2017}.

As tracers of medium and large PAH cations, respectively, the intensities of the 6.2 and 7.7~\mum AIBs and of the 8.6 and 11.0~\mum AIBs can thus be utilized to probe the spatial distributions of these two populations of charged PAHs. We refer to Schefter et al. (2025, in prep.) where the efficacy of the 6.2/8.6 and 7.7/8.6 ratios as measures of the size of the emitting PAHs is explored via their linear relations with the typical PAH size-tracing 3.3/11.2 ratio.

\subsection{The 6.2~\mum AIB as a reliable tracer for PAH ions}
\label{discussion_02}

The population of ionic PAHs in an astronomical environment can be probed by the AIBs at 6.2, 7.7, 8.6, and 11.0~\mum. 
In their study of the PAH emission characteristics in NGC~2023,~\cite{peeters2017} posited that by virtue of their similar behavior on spatial scales, the 8.6 and 11.0~\mum bands serve as equally good tracers of PAH charge, whereas the 7.7~\mum AIB has been shown to be  "contaminated", that is, it comprises two PAH subpopulations: PAH cations, and larger-sized species, for example, PAH clusters~\citep{peeters2017}. However, neither the 8.6 nor the 11.0~\mum band trace PAH charge exclusively, as both bands are also influenced by PAH molecular structure and size. As explored in Sect.~\ref{discussion_01}, both bands are biased toward large, compact PAH cations, and the 11.0~\mum band in particular can also arise from nitrogen-substituted PAHs~\citep{Ricca2021}, making it an ambiguous diagnostic of homocyclic PAH ionization. 

Turning to the 6.2~\mum band, PAH cations exhibit the 6.2~\mum band strongly~\citep{Bauschlicher2009} and~\cite{Maragkoudakis2022} have successfully calibrated the 6.2/11.2~\mum PAH band strength ratio against the PAH ionization parameter~$\gamma$ for a large number of galaxies. However, this band too carries intrinsic biases. Theoretical spectra PAHs show that the intrinsic strengths of the C-C and C-H modes change differently with PAH size~\citep{Bauschlicher:2008, ricca2012, Lemmens2023}.
Particularly, the fraction of absorbed FUV energy emitted in the 6.2~\mum band decreases with increasing PAH size, reflecting an intrinsic strength-driven bias~\citep{Bauschlicher:2008}. In addition, structural family matters: per C atom, the 6.2~\mum band is up to $~\sim$50$\%$ stronger in coronene PAH family than the ovalene family~\citep{Bauschlicher:2008}.

While none of the AIBs provide a bias-free measure of PAH ionization, we conclude that the 6.2/11.2 AIB ratio offers the most representative diagnostic of charged PAHs in the ISM. This is consistent with the recent study by Maragkoudakis et al. 2025b, where utilizing PAHdb and pyPAHdb spectral modeling to analyze the MIRI-MRS observations of the Orion Bar, the authors find the 6.2/11.2 PAH ratio to correlate best with~$\gamma$ than the 7.7/11.2 and 8.6/11.2 ratios. It would therefore be advantageous for future space missions to include a photometric filter centered on 6.2~\mum to capture this feature and probe the ionic population of PAHs in the interstellar medium.

\subsection{JWST MIRI Imaging filters as tracers for the PAH ionization fraction}
\label{discussion_03}

Photometry with JWST has been used for investigating PAH characteristics in a myriad of astronomical environments. Recently, \cite{SEP4} has provided empirical prescriptions for tracing line and PAH emission in the Orion Bar, based on NIRSpec IFU spectroscopy and MIRI images, while \cite{Donnelly2025} have provided a prescription for estimating the PAH 7.7~\mum and 11.2~\mum flux in star-forming regions in four nearby luminous infrared galaxies (LIRGs). We studied a variety of photometry-based prescriptions proposed by \cite{SEP4} to ascertain which would be most effective at tracing the ionized fraction of the PAH population. Since MIRI does not have a filter at 6.2~\mum, we have focused on the viability of the 7.7~\mum filter and its continuum filters as a tracer of PAH ionization. 
We utilize the synthetic images for the MIRI imaging filters F770W, F1000W, F1130W, and F1500W, generated using the IFU spectroscopic data and the linear combinations of these filters for tracing the 7.7~\mum and 11.2~\mum PAH emissions, tested by~\cite{SEP4}. We emphasize that we adopt the global continuum when measuring the 7.7~\mum emission for this purpose, as this approach is more consistent with both the F770W filter response and the continuum used by~\cite{SEP4}. We provide all the proposed prescriptions using these filters in Appendix~\ref{asubsec:prescriptions}, and comment on the main trends as follows.

As seen in Fig.~\ref{fig:photom_spec}, the ratio of the synthetic F770W and F1130W images reproduces well, through a linear relationship, the relative 7.7/11.2 PAH emission. A good match is also obtained when the F1000W and F1500W filters are used to subtract the contribution of the underlying continua from the F770W and F1130W filters tracing PAH 7.7 and 11.2~\mum emission, respectively. However, significant deviations emerge in the \Ta, rooted in the deviations noted by ~\cite{SEP4} when estimating the PAH 7.7~\mum (resp. 11.2~\mum) surface brightnesses using a linear fit of the F770W and F1000W (resp. F1130W and F1000 or F1500W) filters. Indeed, the correlations in Figs.~\ref{fig:photom_spec} and ~\ref{fig:photom_spec_all} involving the continuum filters reveal that the results for the \Ta, which comprises emission from the background face-on PDR, vs. the edge-on PDR are well-separated, where the 7.7/11.2 ratio experiences little change for the face-on PDR. The aforementioned prescriptions thus reveal the effect of face-on vs. edge-on geometry in studying the PAH charge ratio in a target. Positive, linear relations between the photometric filter combinations and the 7.7/11.2 spectroscopic observations exist for the edge-on PDR while the PAH charge ratio as traced by the 7.7/11.2 ratio remains relatively constant for the background face-on PDR. We note, however, that the deviations seen for the \Ta emission disappear when only the F770W/F1130W ratio is considered. We stress that these discrepancies may be specific to the Orion Bar, where the 7.7/11.2 brightness ratio covers only a narrow range. In contrast, studies of larger galaxy samples, covering a broader range in this ratio, may still reveal underlying correlations.

\bigskip

\section{Conclusions}
\label{sec:conclusion}

JWST observations of the Orion Bar have allowed for in-depth investigations of the PAH families prevalent in this archetypal PDR. This study focuses on exploring the efficacy of the AIBs within the 6--9~\mum wavelength domain to probe the charge states and the sizes of the PAHs responsible for these spectral signatures. Analysis of spectroscopic data from the JWST MIRI-MRS IFU instrument yields the following conclusions:

\begin{enumerate}
    \item Based on their similar spatial distributions and correlated surface brightnesses, the charge-diagnostic AIBs at 6.2, 7.7, 8.6, and 11.0~\mum separate into two groups: (i) the 6.2 and 7.7~\mum feature, and (ii) the 8.6 and 11.0~\mum feature. 
    \item The 8.6 and 11.0~\mum bands are characteristic of large, compact, cationic PAHs, whereas the 6.2 and 7.7~\mum bands dominate in large and medium-sized, cationic PAHs. Consequently, the 6.2/8.6 and 7.7/8.6 ratios provide valuable probes of the size distribution of the emitting PAH population, as an alternative to the commonly used 3.3/11.2 size diagnostic. 
    \item As tracers of PAH charge, all AIBs (6.2, 7.7, 8.6, and 11.0~\mum) are inherently biased toward the specific PAH sub-populations they probe. Among them, the 6.2~\mum AIB emerges as the most reliable diagnostic of PAH charge in the ISM.
    \item PAH properties can be studied well through JWST photometry. Myriad prescriptions that utilize MIRI imaging filters at F770W, F1000W, F1130W, and F1500W are proven effective at estimating the PAH ionization fraction, as probed by the 7.7/11.2 PAH emission, in edge-on PDRs. In contrast, the simpler F770W/F1130W photometric ratio provides the best diagnostic for this fraction in face-on PDRs. Since these conclusions are based on the study of the Orion Bar, further studies of additional PDRs are required to probe the full diversity of the astronomical 7.7/11.2 PAH ratio.
\end{enumerate}

\begin{acknowledgements}
This work is based on observations made with the NASA/ESA/CSA James Webb Space Telescope. The data were obtained from the Mikulski Archive for Space Telescopes at the Space Telescope Science Institute, which is operated by the Association of Universities for Research in Astronomy, Inc., under NASA contract NAS 5-03127 for JWST. These observations are associated with program \#1288 (DOI: 10.17909/pg4c-1737).
Support for program \#1288 was provided by NASA through a grant from the Space Telescope Science Institute, which is operated by the Association of Universities for Research in Astronomy, Inc., under NASA contract NAS 5-03127.

Els Peeters and Jan Cami acknowledge support from the University of Western Ontario, the Canadian Space Agency (CSA, 22JWGO1-16), and the Natural Sciences and Engineering Research Council of Canada. 
This article is based upon work from COST Action CA21126 - Carbon molecular nanostructures in space (NanoSpace), supported by COST (European Cooperation in Science and Technology). 
Takashi Onaka acknowledges the support by the Japan Society for the Promotion of Science (JSPS) KAKENHI Grant Number JP24K07087. 
Christiaan Boersma is grateful for an appointment at NASA Ames Research Center through the San Jos\'e State University Research Foundation (80NSSC22M0107).
\end{acknowledgements}
\bibliographystyle{aa}
\bibliography{bib.bib}

\begin{thebibliography}{40}
\expandafter\ifx\csname natexlab\endcsname\relax\def\natexlab#1{#1}\fi

\bibitem[{{Allamandola} {et~al.}(1999){Allamandola}, {Hudgins}, \& {Sandford}}]{Allamandola1999}
{Allamandola}, L.~J., {Hudgins}, D.~M., \& {Sandford}, S.~A. 1999, \apjl, 511, L115

\bibitem[{{Allamandola} {et~al.}(1985){Allamandola}, {Tielens}, \& {Barker}}]{Allamandola1985}
{Allamandola}, L.~J., {Tielens}, A.~G.~G.~M., \& {Barker}, J.~R. 1985, \apjl, 290, L25

\bibitem[{{Allamandola} {et~al.}(1989){Allamandola}, {Tielens}, \& {Barker}}]{Allamandola1989}
{Allamandola}, L.~J., {Tielens}, A.~G.~G.~M., \& {Barker}, J.~R. 1989, \apjs, 71, 733

\bibitem[{{Bally}(2008)}]{Bally2008}
{Bally}, J. 2008, in Handbook of Star Forming Regions, Volume I, ed. B.~{Reipurth}, Vol.~4, 459

\bibitem[{{Bauschlicher} {et~al.}(2008){Bauschlicher}, {Peeters}, \& {Allamandola}}]{Bauschlicher:2008}
{Bauschlicher}, Charles~W., J., {Peeters}, E., \& {Allamandola}, L.~J. 2008, \apj, 678, 316

\bibitem[{{Bauschlicher} {et~al.}(2009){Bauschlicher}, {Peeters}, \& {Allamandola}}]{Bauschlicher2009}
{Bauschlicher}, Charles~W., J., {Peeters}, E., \& {Allamandola}, L.~J. 2009, \apj, 697, 311

\bibitem[{{Bern{\'e}} {et~al.}(2022){Bern{\'e}}, {Habart}, {Peeters}, {Abergel}, {Bergin}, {Bernard-Salas}, {Bron}, {Cami}, {Dartois}, {Fuente}, {Goicoechea}, {Gordon}, {Okada}, {Onaka}, {Robberto}, {R{\"o}llig}, {Tielens}, {Vicente}, {Wolfire}, {Alarc{\'o}n}, {Boersma}, {Canin}, {Chown}, {Dicken}, {Languignon}, {Le Gal}, {Pound}, {Trahin}, {Simmer}, {Sidhu}, {Van De Putte}, {Cuadrado}, {Guilloteau}, {Maragkoudakis}, {Schefter}, {Schirmer}, {Cazaux}, {Aleman}, {Allamandola}, {Auchettl}, {Baratta}, {Bejaoui}, {Bera}, {Bilalbegovi{\'c}}, {Black}, {Boulanger}, {Bouwman}, {Brandl}, {Brechignac}, {Br{\"u}nken}, {Burkhardt}, {Candian}, {Cernicharo}, {Chabot}, {Chakraborty}, {Champion}, {Colgan}, {Cooke}, {Coutens}, {Cox}, {Demyk}, {Donovan Meyer}, {Engrand}, {Foschino}, {Garc{\'\i}a-Lario}, {Gavilan}, {Gerin}, {Godard}, {Gottlieb}, {Guillard}, {Gusdorf}, {Hartigan}, {He}, {Herbst}, {Hornekaer}, {J{\"a}ger}, {Janot-Pacheco}, {Joblin}, {Kaufman}, {Kemper}, {Kendrew}, {Kirsanova}, {Klaassen}, {Knight}, {Kwok},
  {Labiano}, {Lai}, {Lee}, {Lefloch}, {Le Petit}, {Li}, {Linz}, {Mackie}, {Madden}, {Mascetti}, {McGuire}, {Merino}, {Micelotta}, {Misselt}, {Morse}, {Mulas}, {Neelamkodan}, {Ohsawa}, {Omont}, {Paladini}, {Palumbo}, {Pathak}, {Pendleton}, {Petrignani}, {Pino}, {Puga}, {Rangwala}, {Rapacioli}, {Ricca}, {Roman-Duval}, {Roser}, {Roueff}, {Rouill{\'e}}, {Salama}, {Sales}, {Sandstrom}, {Sarre}, {Sciamma-O'Brien}, {Sellgren}, {Shannon}, {Shenoy}, {Teyssier}, {Thomas}, {Togi}, {Verstraete}, {Witt}, {Wootten}, {Ysard}, {Zettergren}, {Zhang}, {Zhang}, \& {Zhen}}]{PDRs4AllPASP}
{Bern{\'e}}, O., {Habart}, {\'E}., {Peeters}, E., {et~al.} 2022, \pasp, 134, 054301

\bibitem[{{Bregman} \& {Temi}(2005)}]{Bregman2005}
{Bregman}, J. \& {Temi}, P. 2005, \apj, 621, 831

\bibitem[{{Candian} {et~al.}(2014){Candian}, {Sarre}, \& {Tielens}}]{Candian2014}
{Candian}, A., {Sarre}, P.~J., \& {Tielens}, A.~G.~G.~M. 2014, \apjl, 791, L10

\bibitem[{{Chown} {et~al.}(2025){Chown}, {Okada}, {Peeters}, {Sidhu}, {Khan}, {Schefter}, {Trahin}, {Canin}, {Van De Putte}, {Alarc{\'o}n}, {Schroetter}, {Kannavou}, {Habart}, {Bern{\'e}}, {Boersma}, {Cami}, {Dartois}, {Goicoechea}, {Gordon}, \& {Onaka}}]{SEP4}
{Chown}, R., {Okada}, Y., {Peeters}, E., {et~al.} 2025, \aap, 698, A86

\bibitem[{{Chown} {et~al.}(2024){Chown}, {Sidhu}, {Peeters}, {Tielens}, {Cami}, {Bern{\'e}}, {Habart}, {Alarc{\'o}n}, {Canin}, {Schroetter}, {Trahin}, {Van De Putte}, {Abergel}, {Bergin}, {Bernard-Salas}, {Boersma}, {Bron}, {Cuadrado}, {Dartois}, {Dicken}, {El-Yajouri}, {Fuente}, {Goicoechea}, {Gordon}, {Issa}, {Joblin}, {Kannavou}, {Khan}, {Lacinbala}, {Languignon}, {Le Gal}, {Maragkoudakis}, {Meshaka}, {Okada}, {Onaka}, {Pasquini}, {Pound}, {Robberto}, {R{\"o}llig}, {Schefter}, {Schirmer}, {Vicente}, {Wolfire}, {Zannese}, {Aleman}, {Allamandola}, {Auchettl}, {Baratta}, {Bejaoui}, {Bera}, {Black}, {Boulanger}, {Bouwman}, {Brandl}, {Brechignac}, {Br{\"u}nken}, {Buragohain}, {Burkhardt}, {Candian}, {Cazaux}, {Cernicharo}, {Chabot}, {Chakraborty}, {Champion}, {Colgan}, {Cooke}, {Coutens}, {Cox}, {Demyk}, {Meyer}, {Foschino}, {Garc{\'\i}a-Lario}, {Gavilan}, {Gerin}, {Gottlieb}, {Guillard}, {Gusdorf}, {Hartigan}, {He}, {Herbst}, {Hornekaer}, {J{\"a}ger}, {Janot-Pacheco}, {Kaufman}, {Kemper}, {Kendrew},
  {Kirsanova}, {Klaassen}, {Kwok}, {Labiano}, {Lai}, {Lee}, {Lefloch}, {Le Petit}, {Li}, {Linz}, {Mackie}, {Madden}, {Mascetti}, {McGuire}, {Merino}, {Micelotta}, {Misselt}, {Morse}, {Mulas}, {Neelamkodan}, {Ohsawa}, {Omont}, {Paladini}, {Palumbo}, {Pathak}, {Pendleton}, {Petrignani}, {Pino}, {Puga}, {Rangwala}, {Rapacioli}, {Ricca}, {Roman-Duval}, {Roser}, {Roueff}, {Rouill{\'e}}, {Salama}, {Sales}, {Sandstrom}, {Sarre}, {Sciamma-O'Brien}, {Sellgren}, {Shenoy}, {Teyssier}, {Thomas}, {Togi}, {Verstraete}, {Witt}, {Wootten}, {Zettergren}, {Zhang}, {Zhang}, \& {Zhen}}]{OBpahs}
{Chown}, R., {Sidhu}, A., {Peeters}, E., {et~al.} 2024, \aap, 685, A75

\bibitem[{{Croiset} {et~al.}(2016){Croiset}, {Candian}, {Bern{\'e}}, \& {Tielens}}]{Croiset2016}
{Croiset}, B.~A., {Candian}, A., {Bern{\'e}}, O., \& {Tielens}, A.~G.~G.~M. 2016, \aap, 590, A26

\bibitem[{{Donnelly} {et~al.}(2025){Donnelly}, {Lai}, {Armus}, {D{\'\i}az-Santos}, {Larson}, {Barcos-Mu{\~n}oz}, {Bianchin}, {Bohn}, {B{\"o}ker}, {Buiten}, {Charmandaris}, {Evans}, {Howell}, {Inami}, {Kakkad}, {Lenki{\'c}}, {Linden}, {Lofaro}, {Malkan}, {Medling}, {Privon}, {Ricci}, {Smith}, {Song}, {Stierwalt}, {van der Werf}, \& {U}}]{Donnelly2025}
{Donnelly}, G.~P., {Lai}, T. S.~Y., {Armus}, L., {et~al.} 2025, \apj, 983, 79

\bibitem[{{Galliano} {et~al.}(2008){Galliano}, {Madden}, {Tielens}, {Peeters}, \& {Jones}}]{Galliano2008}
{Galliano}, F., {Madden}, S.~C., {Tielens}, A. G.~G.~M., {Peeters}, E., \& {Jones}, A.~P. 2008, \apj, 679, 310

\bibitem[{{Gillett} {et~al.}(1973){Gillett}, {Forrest}, \& {Merrill}}]{Gillett1973}
{Gillett}, F.~C., {Forrest}, W.~J., \& {Merrill}, K.~M. 1973, \apj, 183, 87

\bibitem[{{Goicoechea} {et~al.}(2025){Goicoechea}, {Pety}, {Cuadrado}, {Bern{\'e}}, {Dartois}, {Gerin}, {Joblin}, {K{\l}os}, {Lique}, {Onaka}, {Peeters}, {Tielens}, {Alarc{\'o}n}, {Bron}, {Cami}, {Canin}, {Chapillon}, {Chown}, {Fuente}, {Habart}, {Kannavou}, {Le Petit}, {Santa-Maria}, {Schroetter}, {Sidhu}, {Trahin}, {Van De Putte}, \& {Zannese}}]{OBhydrocarbon}
{Goicoechea}, J.~R., {Pety}, J., {Cuadrado}, S., {et~al.} 2025, \aap, 696, A100

\bibitem[{{Habart} {et~al.}(2024){Habart}, {Peeters}, {Bern{\'e}}, {Trahin}, {Canin}, {Chown}, {Sidhu}, {Van De Putte}, {Alarc{\'o}n}, {Schroetter}, {Dartois}, {Vicente}, {Abergel}, {Bergin}, {Bernard-Salas}, {Boersma}, {Bron}, {Cami}, {Cuadrado}, {Dicken}, {Elyajouri}, {Fuente}, {Goicoechea}, {Gordon}, {Issa}, {Joblin}, {Kannavou}, {Khan}, {Lacinbala}, {Languignon}, {Le Gal}, {Maragkoudakis}, {Meshaka}, {Okada}, {Onaka}, {Pasquini}, {Pound}, {Robberto}, {R{\"o}llig}, {Schefter}, {Schirmer}, {Tabone}, {Tielens}, {Wolfire}, {Zannese}, {Ysard}, {Miville-Deschenes}, {Aleman}, {Allamandola}, {Auchettl}, {Baratta}, {Bejaoui}, {Bera}, {Black}, {Boulanger}, {Bouwman}, {Brandl}, {Brechignac}, {Br{\"u}nken}, {Buragohain}, {Burkhardt}, {Candian}, {Cazaux}, {Cernicharo}, {Chabot}, {Chakraborty}, {Champion}, {Colgan}, {Cooke}, {Coutens}, {Cox}, {Demyk}, {Meyer}, {Foschino}, {Garc{\'\i}a-Lario}, {Gavilan}, {Gerin}, {Gottlieb}, {Guillard}, {Gusdorf}, {Hartigan}, {He}, {Herbst}, {Hornekaer}, {J{\"a}ger}, {Janot-Pacheco},
  {Kaufman}, {Kemper}, {Kendrew}, {Kirsanova}, {Klaassen}, {Kwok}, {Labiano}, {Lai}, {Lee}, {Lefloch}, {Le Petit}, {Li}, {Linz}, {Mackie}, {Madden}, {Mascetti}, {McGuire}, {Merino}, {Micelotta}, {Misselt}, {Morse}, {Mulas}, {Neelamkodan}, {Ohsawa}, {Omont}, {Paladini}, {Palumbo}, {Pathak}, {Pendleton}, {Petrignani}, {Pino}, {Puga}, {Rangwala}, {Rapacioli}, {Ricca}, {Roman-Duval}, {Roser}, {Roueff}, {Rouill{\'e}}, {Salama}, {Sales}, {Sandstrom}, {Sarre}, {Sciamma-O'Brien}, {Sellgren}, {Shenoy}, {Teyssier}, {Thomas}, {Togi}, {Verstraete}, {Witt}, {Wootten}, {Zettergren}, {Zhang}, {Zhang}, \& {Zhen}}]{OBim}
{Habart}, E., {Peeters}, E., {Bern{\'e}}, O., {et~al.} 2024, \aap, 685, A73

\bibitem[{{Habing}(1968)}]{Habing1968}
{Habing}, H.~J. 1968, \bain, 19, 421

\bibitem[{{Hony} {et~al.}(2001){Hony}, {Van Kerckhoven}, {Peeters}, {Tielens}, {Hudgins}, \& {Allamandola}}]{hony2001}
{Hony}, S., {Van Kerckhoven}, C., {Peeters}, E., {et~al.} 2001, \aap, 370, 1030

\bibitem[{{Khan} {et~al.}(2025){Khan}, {Abbott, Benjamin}, {Peeters, Els}, {Tielens, Alexander G. G. M.}, {Onaka, Takashi}, {Cami, Jan}, {Schefter, Bethany}, {Boersma, Christiaan}, {Dartois, Emmanuel}, {Goicoechea, Javier R.}, {Maragkoudakis, Alexandros}, {Van De Putte, Dries}, {Buragohain, Mridusmita}, {Candian, Alessandra}, {Labiano, Álvaro}, {Lai, Thomas S.-Y.}, {Ricca, Alessandra}, {Sales, Dinalva A.}, {Zhang, Yong}, {Sidhu, Ameek}, {Chown, Ryan}, {Canin, Amélie}, {Trahin, Boris}, {Schroetter, Ilane}, {Kannavou, Olga}, {Alarcón, Felipe}, {Berné, Olivier}, \& {Habart, Emilie}}]{OBOOPs}
{Khan}, B., {Abbott, Benjamin}, {Peeters, Els}, {et~al.} 2025, A\&A, 699, A133

\bibitem[{{Lemmens} {et~al.}(2023){Lemmens}, {Mackie}, {Candian}, {Lee}, {Tielens}, {Rijs}, \& {Buma}}]{Lemmens2023}
{Lemmens}, A.~K., {Mackie}, C.~J., {Candian}, A., {et~al.} 2023, Faraday Discussions, 245, 380

\bibitem[{{Li}(2020)}]{Li2020}
{Li}, A. 2020, Nature Astronomy, 4, 339

\bibitem[{{Maragkoudakis} {et~al.}(2022){Maragkoudakis}, {Boersma}, {Temi}, {Bregman}, \& {Allamandola}}]{Maragkoudakis2022}
{Maragkoudakis}, A., {Boersma}, C., {Temi}, P., {Bregman}, J.~D., \& {Allamandola}, L.~J. 2022, \apj, 931, 38

\bibitem[{{Maragkoudakis} {et~al.}(2020){Maragkoudakis}, {Peeters}, \& {Ricca}}]{Maragkoudakis2020}
{Maragkoudakis}, A., {Peeters}, E., \& {Ricca}, A. 2020, \mnras, 494, 642

\bibitem[{{Merrill} {et~al.}(1975){Merrill}, {Soifer}, \& {Russell}}]{Merrill1975}
{Merrill}, K.~M., {Soifer}, B.~T., \& {Russell}, R.~W. 1975, \apjl, 200, L37

\bibitem[{{Oomens} {et~al.}(2003){Oomens}, {Tielens}, {Sartakov}, {von Helden}, \& {Meijer}}]{Oomens2003}
{Oomens}, J., {Tielens}, A.~G.~G.~M., {Sartakov}, B.~G., {von Helden}, G., \& {Meijer}, G. 2003, \apj, 591, 968

\bibitem[{{Pathak} \& {Rastogi}(2008)}]{Pathak2008}
{Pathak}, A. \& {Rastogi}, S. 2008, \aap, 485, 735

\bibitem[{{Peeters} {et~al.}(2017){Peeters}, {Bauschlicher}, {Allamandola}, {Tielens}, {Ricca}, \& {Wolfire}}]{peeters2017}
{Peeters}, E., {Bauschlicher}, Charles~W., J., {Allamandola}, L.~J., {et~al.} 2017, \apj, 836, 198

\bibitem[{{Peeters} {et~al.}(2024){Peeters}, {Habart}, {Bern{\'e}}, {Sidhu}, {Chown}, {Van De Putte}, {Trahin}, {Schroetter}, {Canin}, {Alarc{\'o}n}, {Schefter}, {Khan}, {Pasquini}, {Tielens}, {Wolfire}, {Dartois}, {Goicoechea}, {Maragkoudakis}, {Onaka}, {Pound}, {Vicente}, {Abergel}, {Bergin}, {Bernard-Salas}, {Boersma}, {Bron}, {Cami}, {Cuadrado}, {Dicken}, {Elyajouri}, {Fuente}, {Gordon}, {Issa}, {Joblin}, {Kannavou}, {Lacinbala}, {Languignon}, {Le Gal}, {Meshaka}, {Okada}, {Robberto}, {R{\"o}llig}, {Schirmer}, {Tabone}, {Zannese}, {Aleman}, {Allamandola}, {Auchettl}, {Baratta}, {Bejaoui}, {Bera}, {Black}, {Boulanger}, {Bouwman}, {Brandl}, {Brechignac}, {Br{\"u}nken}, {Buragohain}, {Burkhardt}, {Candian}, {Cazaux}, {Cernicharo}, {Chabot}, {Chakraborty}, {Champion}, {Colgan}, {Cooke}, {Coutens}, {Cox}, {Demyk}, {Meyer}, {Foschino}, {Garc{\'\i}a-Lario}, {Gerin}, {Gottlieb}, {Guillard}, {Gusdorf}, {Hartigan}, {He}, {Herbst}, {Hornekaer}, {J{\"a}ger}, {Janot-Pacheco}, {Kaufman}, {Kendrew}, {Kirsanova},
  {Klaassen}, {Kwok}, {Labiano}, {Lai}, {Lee}, {Lefloch}, {Le Petit}, {Li}, {Linz}, {Mackie}, {Madden}, {Mascetti}, {McGuire}, {Merino}, {Micelotta}, {Misselt}, {Morse}, {Mulas}, {Neelamkodan}, {Ohsawa}, {Paladini}, {Palumbo}, {Pathak}, {Pendleton}, {Petrignani}, {Pino}, {Puga}, {Rangwala}, {Rapacioli}, {Ricca}, {Roman-Duval}, {Roser}, {Roueff}, {Rouill{\'e}}, {Salama}, {Sales}, {Sandstrom}, {Sarre}, {Sciamma-O'Brien}, {Sellgren}, {Shenoy}, {Teyssier}, {Thomas}, {Togi}, {Verstraete}, {Witt}, {Wootten}, {Ysard}, {Zettergren}, {Zhang}, {Zhang}, \& {Zhen}}]{OBspec}
{Peeters}, E., {Habart}, E., {Bern{\'e}}, O., {et~al.} 2024, \aap, 685, A74

\bibitem[{{Peeters} {et~al.}(2002){Peeters}, {Hony}, {Van Kerckhoven}, {Tielens}, {Allamandola}, {Hudgins}, \& {Bauschlicher}}]{Peeters2002}
{Peeters}, E., {Hony}, S., {Van Kerckhoven}, C., {et~al.} 2002, \aap, 390, 1089

\bibitem[{{Ricca} {et~al.}(2012){Ricca}, {Bauschlicher}, {Boersma}, {Tielens}, \& {Allamandola}}]{ricca2012}
{Ricca}, A., {Bauschlicher}, Charles~W., J., {Boersma}, C., {Tielens}, A. G.~G.~M., \& {Allamandola}, L.~J. 2012, \apj, 754, 75

\bibitem[{{Ricca} {et~al.}(2021){Ricca}, {Boersma}, \& {Peeters}}]{Ricca2021}
{Ricca}, A., {Boersma}, C., \& {Peeters}, E. 2021, \apj, 923, 202

\bibitem[{{Schroetter} {et~al.}(2024){Schroetter}, {Bern{\'e}}, {Joblin}, {Canin}, {Chown}, {Sidhu}, {Habart}, {Peeters}, {Lai}, {Candian}, {Chakraborty}, {Petrignani}, {Trahin}, {Van De Putte}, \& {Alarc{\'o}n}}]{Schroetter2024}
{Schroetter}, I., {Bern{\'e}}, O., {Joblin}, C., {et~al.} 2024, \aap, 685, A78

\bibitem[{{Schutte} {et~al.}(1993){Schutte}, {Tielens}, \& {Allamandola}}]{Schutte1993}
{Schutte}, W.~A., {Tielens}, A.~G.~G.~M., \& {Allamandola}, L.~J. 1993, \apj, 415, 397

\bibitem[{{Shannon} {et~al.}(2016){Shannon}, {Stock}, \& {Peeters}}]{shannon2016}
{Shannon}, M.~J., {Stock}, D.~J., \& {Peeters}, E. 2016, \apj, 824, 111

\bibitem[{{Sidhu} {et~al.}(2021){Sidhu}, {Peeters}, {Cami}, \& {Knight}}]{sidhu2021}
{Sidhu}, A., {Peeters}, E., {Cami}, J., \& {Knight}, C. 2021, \mnras, 500, 177

\bibitem[{{Tielens}(2008)}]{Tielens2008}
{Tielens}, A.~G.~G.~M. 2008, \araa, 46, 289

\bibitem[{{Tielens} \& {Hollenbach}(1985)}]{Tielens1985b}
{Tielens}, A.~G.~G.~M. \& {Hollenbach}, D. 1985, \apj, 291, 747

\bibitem[{{Van De Putte} {et~al.}(2024){Van De Putte}, {Meshaka}, {Trahin}, {Habart}, {Peeters}, {Bern{\'e}}, {Alarc{\'o}n}, {Canin}, {Chown}, {Schroetter}, {Sidhu}, {Boersma}, {Bron}, {Dartois}, {Goicoechea}, {Gordon}, {Onaka}, {Tielens}, {Verstraete}, {Wolfire}, {Abergel}, {Bergin}, {Bernard-Salas}, {Cami}, {Cuadrado}, {Dicken}, {Elyajouri}, {Fuente}, {Joblin}, {Khan}, {Lacinbala}, {Languignon}, {Le Gal}, {Maragkoudakis}, {Okada}, {Pasquini}, {Pound}, {Robberto}, {R{\"o}llig}, {Schefter}, {Schirmer}, {Tabone}, {Vicente}, {Zannese}, {Colgan}, {He}, {Rouill{\'e}}, {Togi}, {Aleman}, {Auchettl}, {Baratta}, {Bejaoui}, {Bera}, {Black}, {Boulanger}, {Bouwman}, {Brandl}, {Brechignac}, {Br{\"u}nken}, {Buragohain}, {Burkhardt}, {Candian}, {Cazaux}, {Cernicharo}, {Chabot}, {Chakraborty}, {Champion}, {Cooke}, {Coutens}, {Cox}, {Demyk}, {Meyer}, {Foschino}, {Garc{\'\i}a-Lario}, {Gerin}, {Gottlieb}, {Guillard}, {Gusdorf}, {Hartigan}, {Herbst}, {Hornekaer}, {Issa}, {J{\"a}ger}, {Janot-Pacheco}, {Kannavou}, {Kaufman},
  {Kemper}, {Kendrew}, {Kirsanova}, {Klaassen}, {Kwok}, {Labiano}, {Lai}, {Le Floch}, {Le Petit}, {Li}, {Linz}, {Mackie}, {Madden}, {Mascetti}, {McGuire}, {Merino}, {Micelotta}, {Morse}, {Mulas}, {Neelamkodan}, {Ohsawa}, {Omont}, {Paladini}, {Palumbo}, {Pathak}, {Pendleton}, {Petrignani}, {Pino}, {Puga}, {Rangwala}, {Rapacioli}, {Rho}, {Ricca}, {Roman-Duval}, {Roser}, {Roueff}, {Salama}, {Sales}, {Sandstrom}, {Sarre}, {Sciamma-O'Brien}, {Sellgren}, {Shenoy}, {Teyssier}, {Thomas}, {Witt}, {Wootten}, {Ysard}, {Zettergren}, {Zhang}, {Zhang}, \& {Zhen}}]{OBlines}
{Van De Putte}, D., {Meshaka}, R., {Trahin}, B., {et~al.} 2024, \aap, 687, A86

\bibitem[{{Wright} {et~al.}(2015){Wright}, {Wright}, {Goodson}, {Rieke}, {Aitink-Kroes}, {Amiaux}, {Aricha-Yanguas}, {Azzollini}, {Banks}, {Barrado-Navascues}, {Belenguer-Davila}, {Blommaert}, {Bouchet}, {Brandl}, {Colina}, {Detre}, {Diaz-Catala}, {Eccleston}, {Friedman}, {Garc{\'\i}a-Mar{\'\i}n}, {G{\"u}del}, {Glasse}, {Glauser}, {Greene}, {Groezinger}, {Grundy}, {Hastings}, {Henning}, {Hofferbert}, {Hunter}, {Jessen}, {Justtanont}, {Karnik}, {Khorrami}, {Krause}, {Labiano}, {Lagage}, {Langer}, {Lemke}, {Lim}, {Lorenzo-Alvarez}, {Mazy}, {McGowan}, {Meixner}, {Morris}, {Morrison}, {M{\"u}ller}, {rgaard-Nielson}, {Olofsson}, {O'Sullivan}, {Pel}, {Penanen}, {Petach}, {Pye}, {Ray}, {Renotte}, {Renouf}, {Ressler}, {Samara-Ratna}, {Scheithauer}, {Schneider}, {Shaughnessy}, {Stevenson}, {Sukhatme}, {Swinyard}, {Sykes}, {Thatcher}, {Tikkanen}, {van Dishoeck}, {Waelkens}, {Walker}, {Wells}, \& {Zhender}}]{MIRI}
{Wright}, G.~S., {Wright}, D., {Goodson}, G.~B., {et~al.} 2015, \pasp, 127, 595

\end{thebibliography}

\begin{appendix}

\section{JWST NIRCam image of the Orion Bar}
\label{FOV}
\begin{figure*}
\centering
\includegraphics[width=0.8\textwidth]{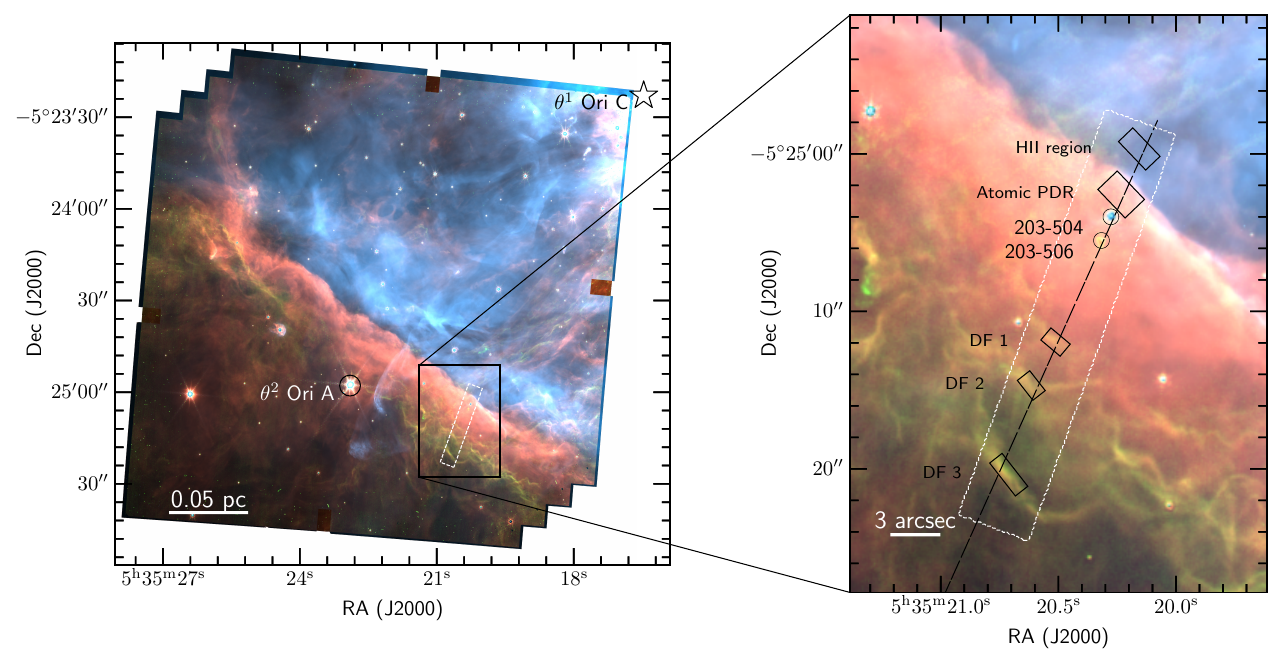}
\caption{A composite JWST NIRCam image of the Orion Bar. The JWST-MIRI/MRS IFU FOV is overlaid in white, and the spectral extraction apertures for the five template spectra are indicated with labels and black boxes in the right panel. The red, green, and blue colors encode the F335M (the 3.3~\mum AIB), F470N-F444W (\molh emission), and F187N (Paschen $\alpha$ emission), respectively \cite{OBim}. The ionizing source for the Orion Bar, \oric,  is represented by the $\star$ symbol on the top right edge of the left panel. In the right panel, the two proplyds 203-504 and 203-506 are shown through black circles, and the black dashed line indicates the cut across the MIRI mosaic (position angle, PA, of 155.79°). This figure is adapted from \cite{OBpahs}, with permission.} 
\label{fig:FOV}
\end{figure*}

\twocolumn
\section{Integration ranges}
\label{asubsec:ranges}

The surface brightnesses of the AIBs studied in this paper were obtained through integrating the specific surface brightnesses as outlined in ~\cite{OBOOPs}. The wavelength ranges used for these integrations are provided in Table~\ref{tab:ranges}.

\begin{table}
\caption{Wavelength ranges for integrating the specific surface brightness of the AIBs studied.}
\label{tab:ranges}
\centering
\begin{tabular}{lll}
\hline\hline
AIB   & $\lambda_\mathrm{min}$ ($\mu$m) & $\lambda_\mathrm{max}$ ($\mu$m) \\
\hline
6.2   & 6.0                             & 6.6  \\
7.7   & 7.05                            & 8.15 \\
8.6   & 8.15                            & 9.1  \\
11.0  & 10.9                            & 11.1 \\
11.2  & 11.1                            & 11.8 \\
\hline
\end{tabular}
\end{table}

\section{Additional Maps, radial profiles and correlation plots}
\label{asubsec:more_maps}

\begin{figure}[h!]
    \begin{center}
   \resizebox{1\columnwidth}{!}{
   \includegraphics[]{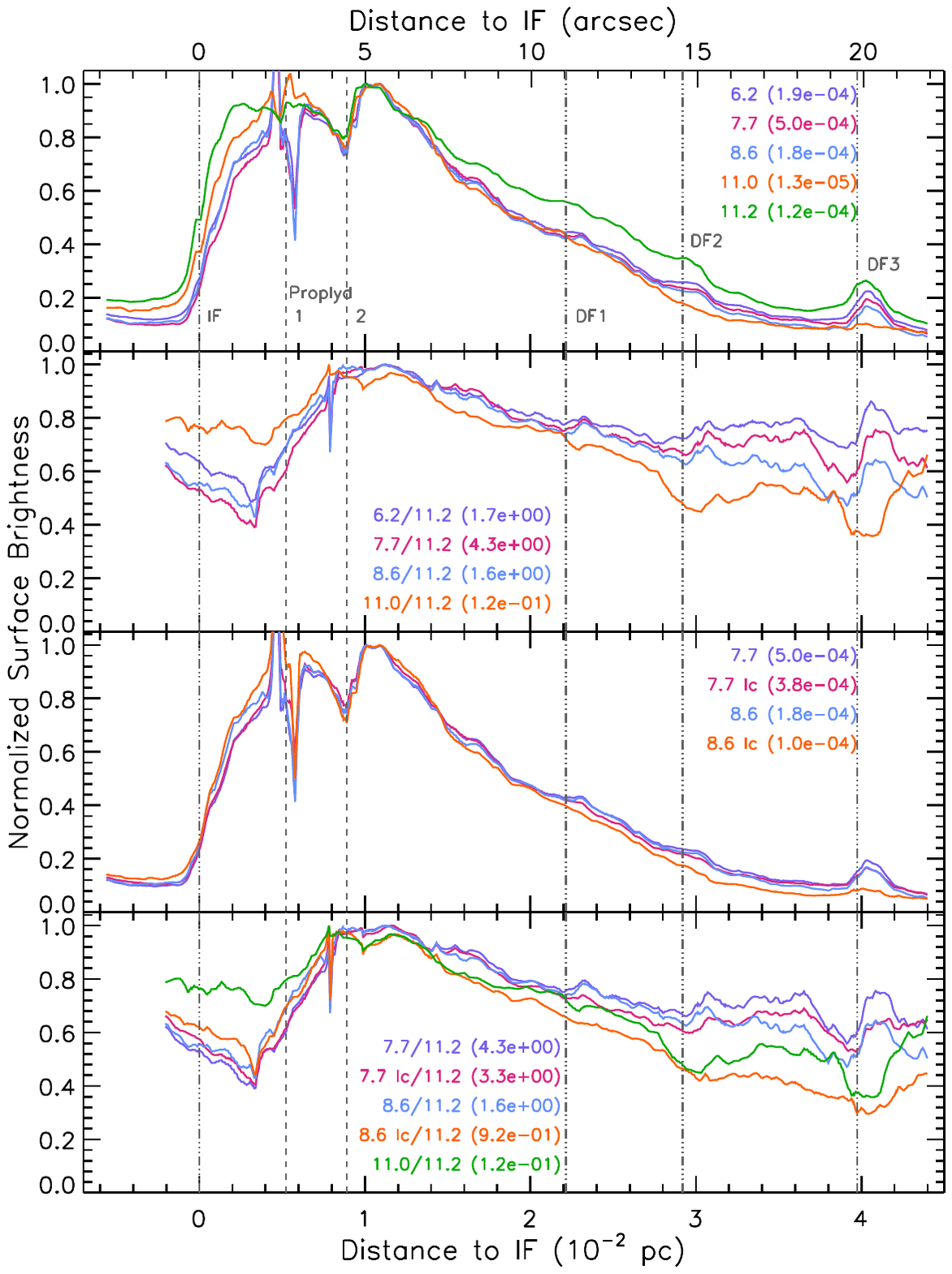}}
   \caption{Normalized surface brightnesses and their ratios for the 6.2, 7.7, 8.6, 11.0, and 11.2~\mum AIBs as a function of distance to the IF (0.228 pc or 113.4\arcsec\, from \oric) along a cut crossing the mosaic (see Fig.~\ref{fig:maps} for the location of the cut). Here, we include radial profies for the 7.7 and 8.6~\mum AIBs measured using both the local and global continua (Fig.~\ref{fig:ADPR_spec}), for comparison. Normalization factors are listed in \usurfacebrightnessalt in parentheses for each surface brightness. As the cut is not perpendicular to the IF and distances are given along the cut, a correction factor of cos(19.58$\degr$)~$=$ 0.942 needs to be applied to obtain a perpendicular distance from the IF.}
   \label{fig:cuts_all}
    \end{center}
\end{figure}

\begin{figure*}[h!]
    \begin{center}
        \resizebox{.9\hsize}{!}{
    \includegraphics{Figures/Maps/map_6p2.pdf}
    \includegraphics{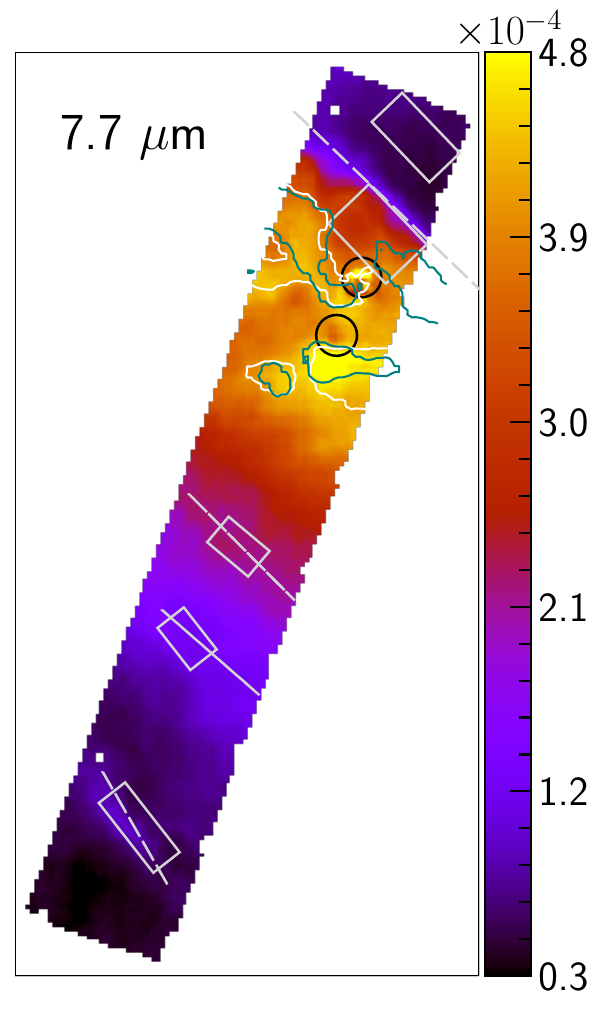}
    \includegraphics{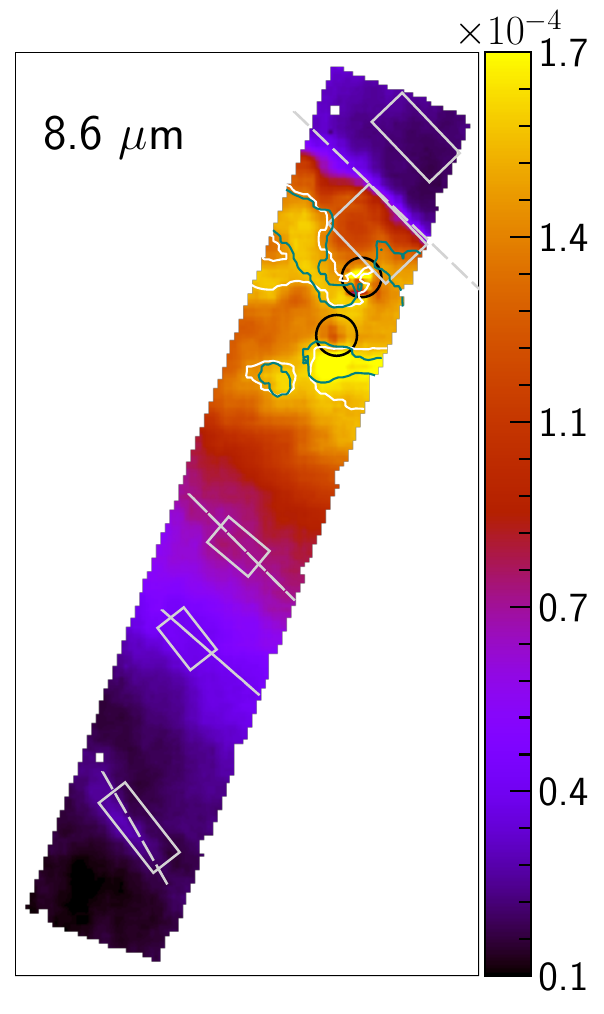}
    \includegraphics{Figures/Maps/map_11p0.pdf}
    \includegraphics{Figures/Maps/map_11p2.pdf}}
        \resizebox{.72\hsize}{!}{
    \includegraphics{Figures/Maps/map_6p2_11p2.pdf} 
    \includegraphics{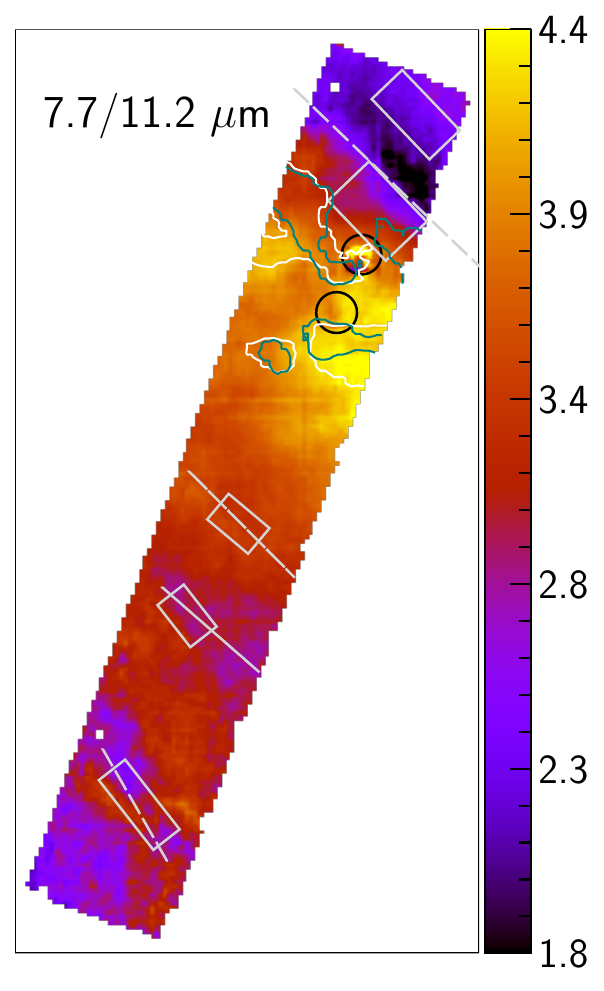}    
    \includegraphics{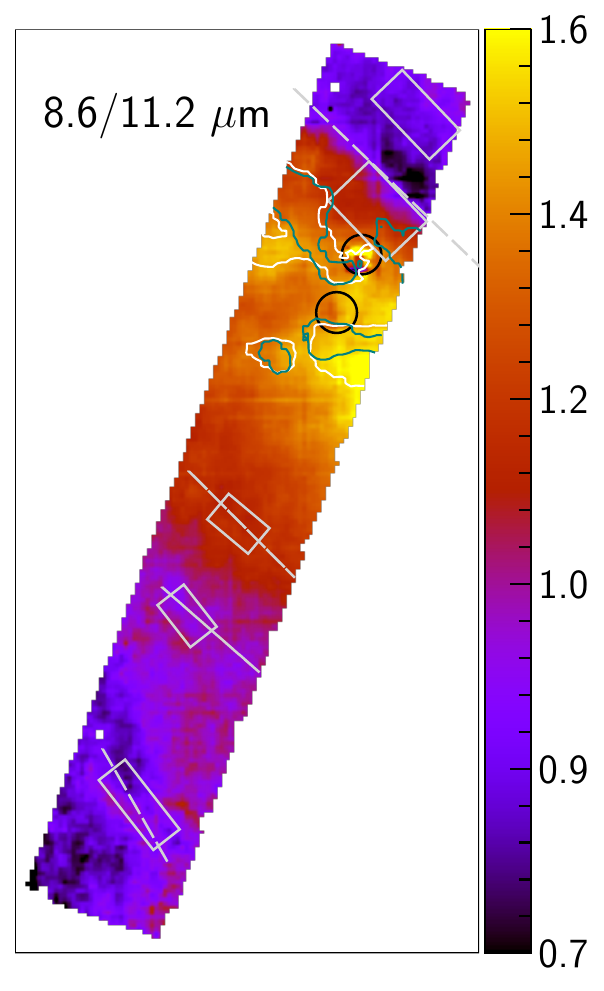}
    \includegraphics{Figures/Maps/map_11p0_11p2.pdf}
    }
     \resizebox{.72\hsize}{!}{
    \includegraphics{Figures/Maps/map_7p7_localcont.pdf} 
    \includegraphics{Figures/Maps/map_8p6_localcont.pdf}    
    \includegraphics{Figures/Maps/map_7p7lc_11p2.pdf}
    \includegraphics{Figures/Maps/map_8p6lc_11p2.pdf}
    }
    \caption{Spatial distribution of the surface brightnesses of the 6.2, 7.7, 8.6, 11.0, and 11.2~\mum AIBs in the Orion Bar PDR, in units of \usurfacebrightnessalt, and brightness ratios relative to the 11.2~\mum AIB. Here, maps of the 7.7 and 8.6~\mum AIBs measured using both the local and global continua (Fig.~\ref{fig:ADPR_spec}) are presented for comparison. 
    \oric is located toward the top right of each map. For each map, the range of the corresponding color bar is set between 0.5$\%$ and 99.5$\%$ percentile level for the data, while zero pixels, edge pixels, and pixels covering the two proplyds as seen in the MIRI mosaic, indicated by the black circles, are masked out. The contours trace peak emission for the 11.0~\mum AIB (white), the 11.2~\mum AIB (teal). The rectangular apertures of the template spectra for the \Ta, \Tb, \Tc, \Td, and \Te, from top to bottom,
    are shown in gray, the gray lines delineate the IF and the three dissociation fronts, \Tc, \Td and \Te, and the dashed, diagonal white line in the top left map indicates the cut across the MIRI mosaic (position angle of 155.79°).}
    \label{fig:maps_all}
    \end{center}
\end{figure*}

\begin{figure*}[ht!]
    \begin{center}
    \begin{tabular}{c}
        \vspace{-.5em}        \includegraphics[width=0.25\textwidth]{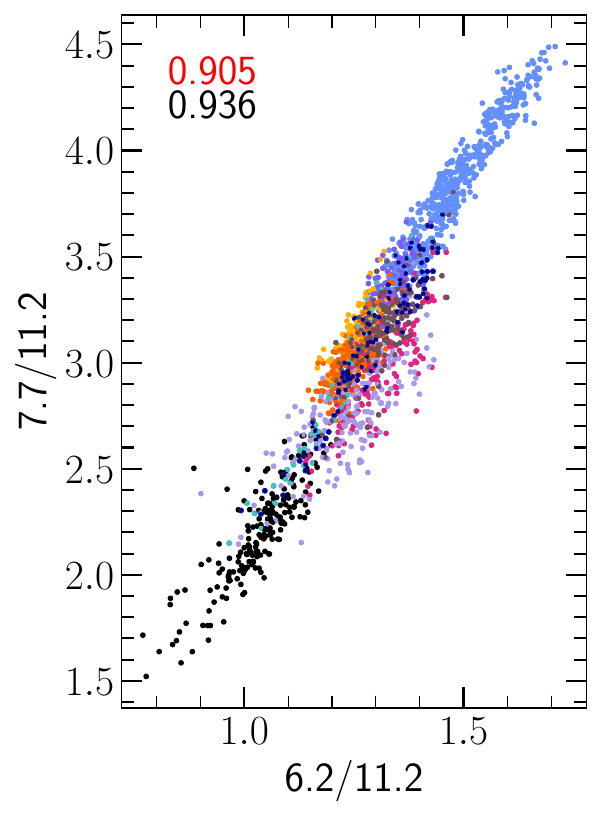}
            \includegraphics[width=0.25\textwidth]{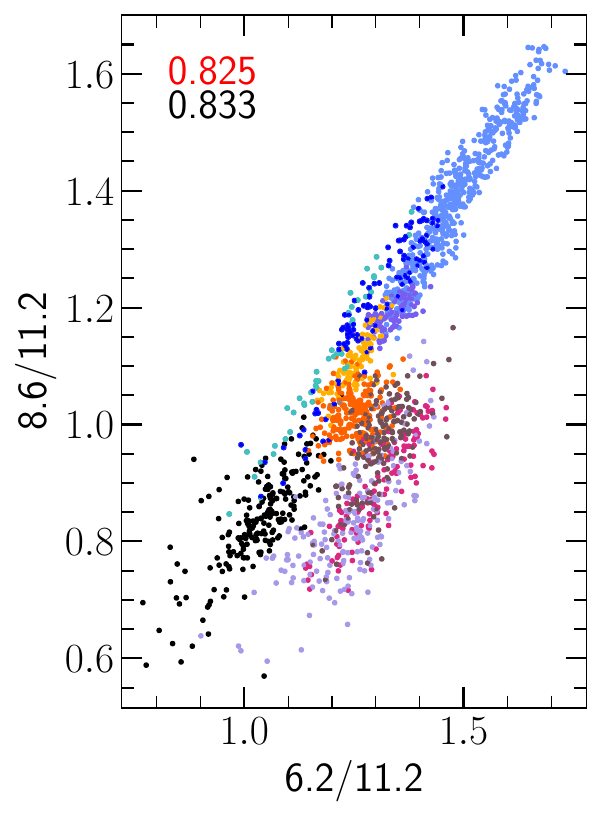}
            \includegraphics[width=0.25\textwidth]{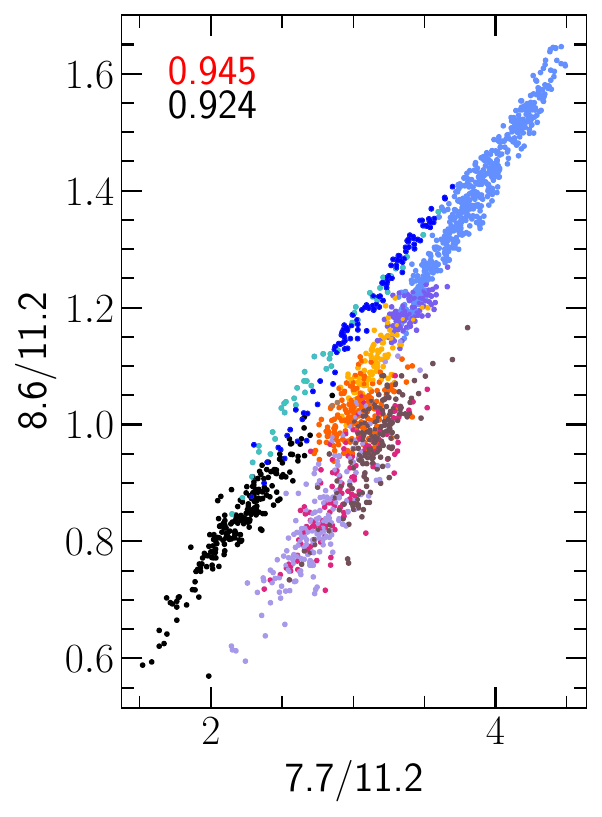}
            \raisebox{.8cm}[0pt][0pt]
            {\includegraphics[height=0.22\textheight]{Figures/Correlations/OB_MIRI_ColorMap_vs2.pdf}}
     \vspace{-.2em} \\
     
            \includegraphics[width=0.25\textwidth]{Figures/Correlations/Corr_Intensities_7p7lc_6p2_over11p2.pdf}
            \includegraphics[width=0.25\textwidth]{Figures/Correlations/Corr_Intensities_8p6lc_6p2_over11p2.pdf}
            \includegraphics[width=0.25\textwidth]{Figures/Correlations/Corr_Intensities_8p6lc_7p7lc_over11p2.pdf}

            \includegraphics[width=0.25\textwidth]{Figures/Correlations/Corr_Intensities_6p2_11p0_over11p2.pdf}
            \vspace{-.5em} \\
             
            \includegraphics[width=0.25\textwidth]{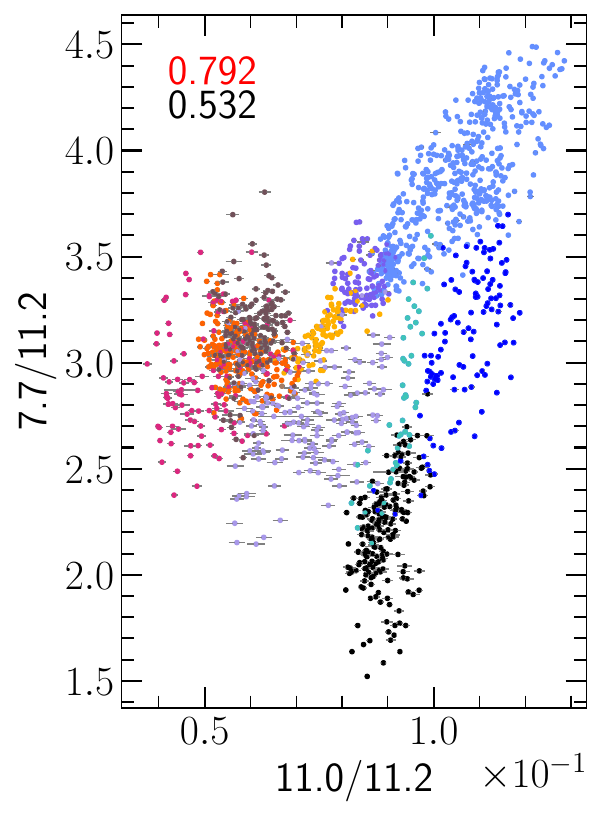}
            \includegraphics[width=0.25\textwidth]{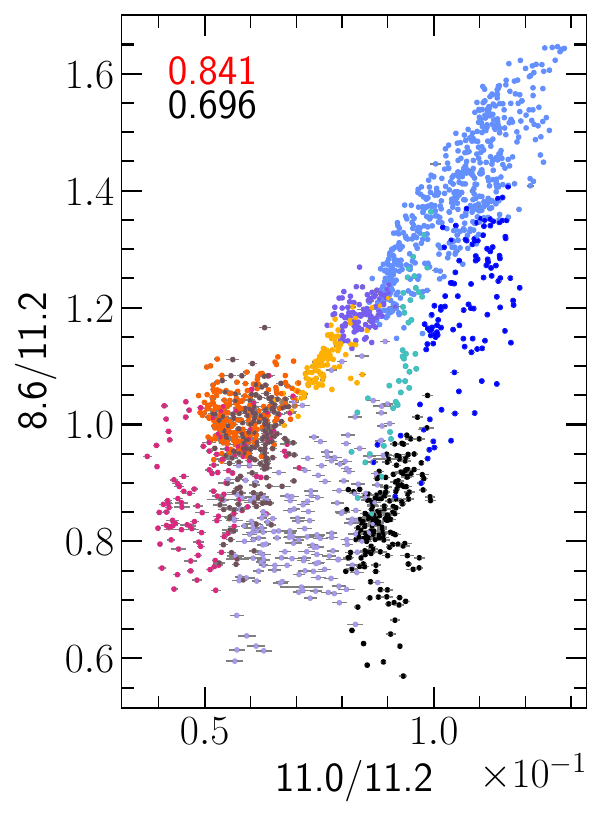}
                        
              \includegraphics[width=0.25\textwidth]{Figures/Correlations/Corr_Intensities_7p7lc_11p0_over11p2.pdf}
              \includegraphics[width=0.25\textwidth]{Figures/Correlations/Corr_Intensities_8p6lc_11p0_over11p2.pdf}
       
    \end{tabular}
 \caption{Correlations of the 7.7/11.2, 8.6/11.2, and 11.0/11.2 AIB surface brightness ratios, where plots involving  the 7.7 and 8.6~\mum AIBs measured using both the local and global continua (Fig.~\ref{fig:ADPR_spec}) are presented for comparison. The Spearman correlation coefficient $\rho$ for the variables excluding
(including) the points from the \HII region (black), and regions beneath the IF (teal and dark blue), are printed in red
(black) in the panels. Only surface brightnesses from every other spaxel are considered in the correlation analyses. The data points are colored according to regions in the mosaic where those pixels are located (top right panel).}
    \label{fig:corr_all}
    \end{center}
\end{figure*}

We present all the radial profiles, spectral maps and correlation plots, including those for the 7.7 and 8.6~\mum AIBs measured using both the local and global continua (Fig.~\ref{fig:ADPR_spec}), in Fig.~\ref{fig:cuts_all},~\ref{fig:maps_all}, and~\ref{fig:corr_all}, respectively. 

There exist subtle differences in morphology at the dissociation fronts when the 7.7 and 8.6~\mum are measured using the two continua, as seen in Fig.~\ref{fig:cuts_all}. The 8.6~\mum surface brightness declines faster across the dissociation fronts when the local continuum is used instead of the global continuum. The 8.6~\mum AIB measured using the global continuum exhibits local peaks just after each dissociation front, which  are no longer detected with the local continuum. This difference in behavior for the 7.7~\mum AIB is minimal. Thus, with the choice of the local continuum, the enhanced emission of the 8.6~\mum in the molecular PDR detected with the global continuum is no longer seen.

Per Fig.~\ref{fig:corr_all}, the branches observed in the 8.6 vs. 7.7 and 6.2~\mum plots are well-separated when taking into account the local continuum instead of the global continuum.
We note that with the choice of the global continuum to measure
the 7.7 and 8.6~\mum bands, the normalized 8.6 vs.11.0~\mum AIB correlation worsens, while the correlations with the 6.2 and 7.7~\mum AIB surface brightnesses improve in terms of the values of R. Overall, the choice of the continuum significantly
influences the correlations involving the 8.6~\mum AIB.

\section{PAH 7.7/11.2 ratio prescriptions}
\label{asubsec:prescriptions}

\begin{figure}[h!]
\begin{center}
\resizebox{.5\columnwidth}{!}{
\includegraphics{Figures/Discussions/photometry/7p7_01_11p2_01_photom_spec.pdf}}
\resizebox{\columnwidth}{!}{\includegraphics{Figures/Discussions/photometry/7p7_03_11p2_04_photom_spec.pdf}
\includegraphics{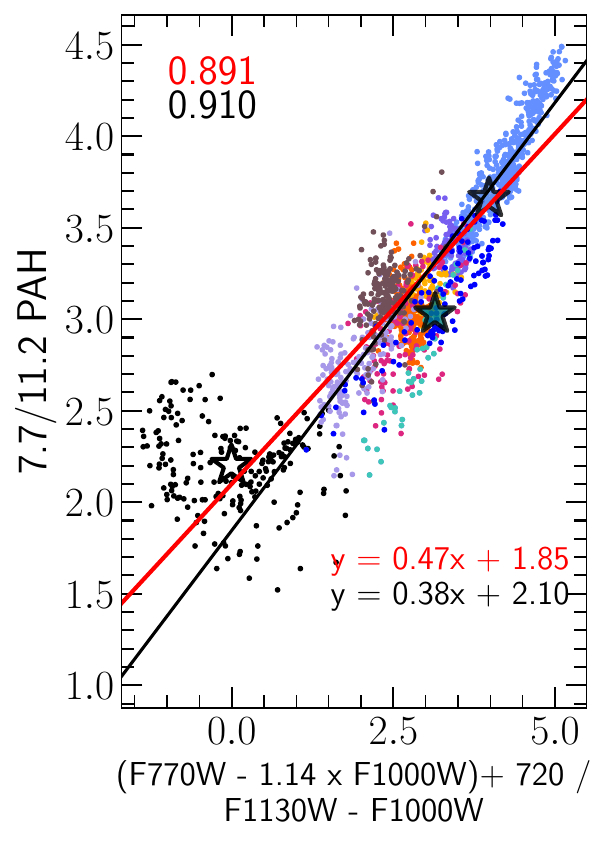}}
\resizebox{\columnwidth}{!}{\includegraphics{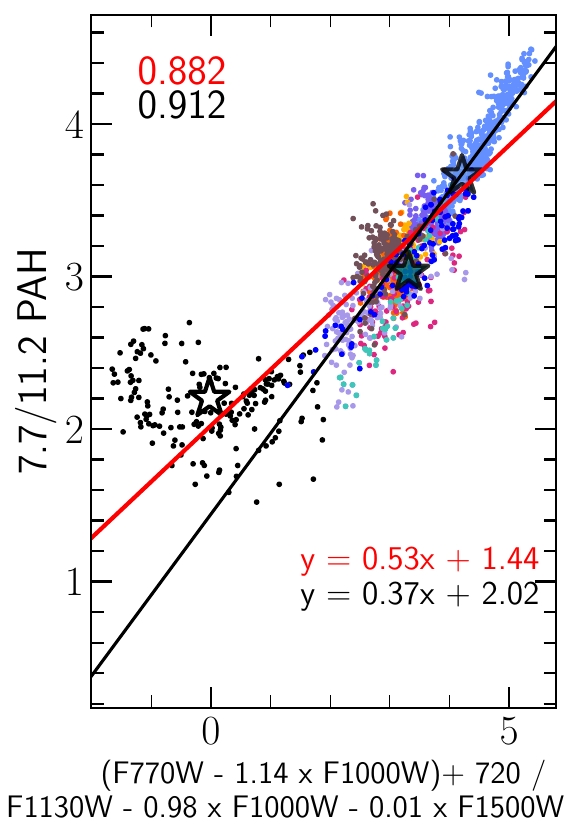}
\includegraphics{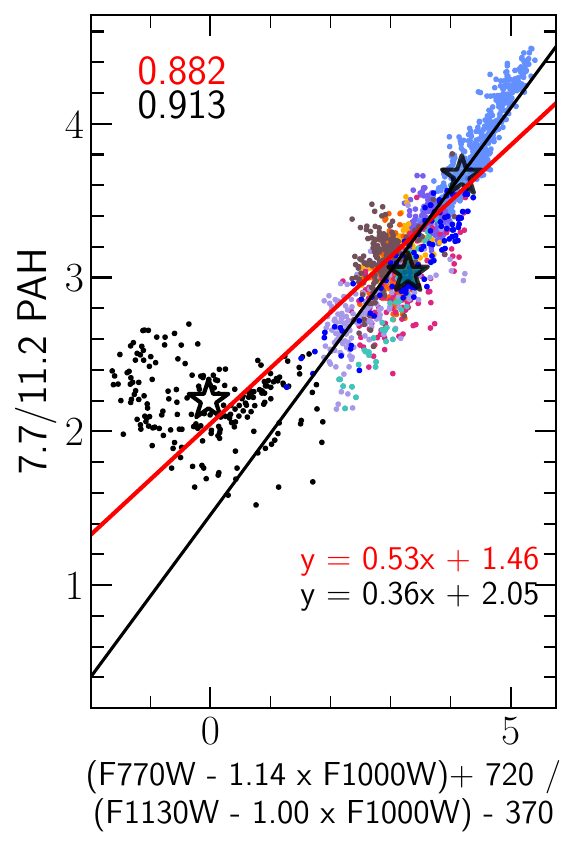}}

\caption{Correlations between ratios of the synthetic images involving F770W, F1130W, and F1500W and the measured 7.7/11.2 spectroscopic PAH emission in the Orion Bar. Here the 7.7~\mum feature is measured using the global continuum (Fig.~\ref{fig:ADPR_spec}). Lines of best-fit through the data excluding
(including) the points from the \Ta (black), and the regions near the IF (teal and dark blue), their equations and the Spearman correlation coefficients R are shown in red (black). The star symbols highlight the average values for the \Tb (blue), the \Ta (black) and the region near the IF (teal and dark blue).}
\label{fig:photom_spec_all}
\end{center}
\end{figure}

The MIRI filter combinations that were tested to match the images to the 7.7/11.2 PAH emission ratio in the Orion bar PDR are presented in Table~\ref{tab:photom_spec}, in decreasing order of the correlation coefficients for the edge-on Orion Bar PDR, which are presented in the correlation plots in Fig.~\ref{fig:photom_spec_all}.

\begin{table}[h!]
\caption{Suggested MIRI filter combinations to recover the PAH 7.7/11.2 ratio.}
\label{tab:photom_spec}
\centering
\begin{tabular}{lll}
\hline\hline
No.    & 7.7 $\mu$m PAH filter      & 11.2 $\mu$m PAH filter          \\
\hline
1      & F770W                      & F1130W                              \\
2      & F770W~$-$ 1.14~$\times$    & F1130W~$-$ 0.29~$\times$            \\
       & F1000W~$+$ 720             & F1500W~$-$ 450                      \\
3      & F770W~$-$ 1.14~$\times$    & F1130W~$-$ F1000W                   \\
       & F1000W~$+$ 720             &                                     \\
4      & F770W~$-$ 1.14~$\times$    & F1130W~$-$ 0.98~$\times$ F1000W~$-$ \\
       & F1000W~$+$ 720             & 0.01~$\times$ F1500W~$-$ 360        \\
5      & F770W~$-$ 1.14~$\times$    & F1130W~$-$ F1000W $-$ 370           \\
       & F1000W~$+$ 720             &                                     \\
\hline
\end{tabular}
\tablefoot{No. denotes the order number defined in the text of Appendix~\ref{asubsec:prescriptions}.}
\end{table}

\end{appendix}

\end{document}